\documentclass[journal, comsoc]{IEEEtran}

\usepackage{ifpdf}
\usepackage{cite}
\usepackage{subfigure}
\ifCLASSINFOpdf
	\usepackage[pdftex]{graphicx, color}
\else
  	\usepackage[dvipdfmx]{graphicx, color}
\fi
\usepackage[cmex10]{amsmath}
\usepackage[caption=false,font=footnotesize]{subfig}

\ifCLASSOPTIONcaptionsoff
	\usepackage[nomarkers]{endfloat}
	\let\MYoriglatexcaption\caption
	\renewcommand{\caption}[2][\relax]{\MYoriglatexcaption[#2]{#2}}
\fi
\usepackage{url}
\usepackage{amssymb}
\usepackage{amsthm}
\usepackage{mathtools}
\usepackage{flushend}
\usepackage{booktabs}
\usepackage{color}
\usepackage{bm,bbm}
\usepackage{ulem}
\usepackage{algorithm,algpseudocode}
\algdef{SE}[DOWHILE]{Do}{DoWhile}{\algorithmicdo}[1]{\algorithmicwhile\ #1}
\algdef{SE}[SUBALG]{Indent}{EndIndent}{}{\algorithmicend\ }%
\algtext*{Indent}
\algtext*{EndIndent}

\usepackage{comment}

\renewcommand{\emph}[1]{\textit{#1}}

\DeclareMathOperator*{\argmax}{arg\,max}

\theoremstyle{definition}

\title{Handover Management for mmWave Networks with Proactive Performance Prediction Using \\ Camera Images and Deep Reinforcement Learning}

\author{
	Yusuke~Koda,~\IEEEmembership{Student~Member,~IEEE,}
	Kota~Nakashima,~\IEEEmembership{Student~Member,~IEEE,}
	Koji~Yamamoto,~\IEEEmembership{Member,~IEEE,}
	Takayuki~Nishio,~\IEEEmembership{Member,~IEEE,}
	and~Masahiro~Morikura,~\IEEEmembership{Member,~IEEE}
	\thanks{
		The authors are with the Graduate School of Informatics, Kyoto University, Yoshida-honmachi, Sakyo-ku, Kyoto 606-8501, Japan, e-mail: \{koda@imc.cce., nakashima@imc.cce, kyamamot@, nishio@, morikura@\}i.kyoto-u.ac.jp.
		This work was supported in part by JSPS KAKENHI Grant Numbers JP17H03266, JP18H01442, and KDDI Foundation.
	}
}
\begin{document}

\maketitle

\begin{abstract}
	For millimeter-wave networks, this paper presents a paradigm shift for leveraging time-consecutive camera images in handover decision problems.
	While making handover decisions, it is important to predict future long-term performance---e.g., the cumulative sum of time-varying data rates---proactively to avoid making myopic decisions.
	However, this study experimentally notices that a time-variation in the received powers is not necessarily informative for proactively predicting the rapid degradation of data rates caused by moving obstacles.
	To overcome this challenge, this study proposes a proactive framework wherein handover timings are optimized while obstacle-caused data rate degradations are predicted before the degradations occur.
	The key idea is to expand a state space to involve time-consecutive camera images, which comprises informative features for predicting such data rate degradations.
	To overcome the difficulty in handling the large dimensionality of the expanded state space, we use a deep reinforcement learning for deciding the handover timings.
	The evaluations performed based on the experimentally obtained camera images and received powers demonstrate that the expanded state space facilitates (i) the  prediction of obstacle-caused data rate degradations from 500\,ms before the degradations occur and (ii) superior performance to a handover framework without the state space expansion.
\end{abstract}

\IEEEpeerreviewmaketitle


\section{Introduction} \label{sec:intro}

\IEEEPARstart{M}{illimeter-wave} (mmWave) communications are expected to play an important role in next-generation wireless networks, such as fifth-generation mobile networks or wireless local area networks\cite{11ad_sakaguchi, millimeter_survey, mmwave_5g, mmwave_5g2}.
The exploitation of wider spectrum bands in the mmWave band facilitates multi-gigabit data transmission and thereby supports communication services, such as ultra-high-definition televisions\cite{millimeter_survey}, virtual reality (VR)\cite{osseiran2014scenarios}, or augmented reality (AR)\cite{series2015imt} that require the multi-gigabit data transmission.

However, designing robust millimeter networks is quite challenging owing to the high frequency of the mmWave bands.
The distinct feature of mmWave communication is the use of directional antennas to compensate for high path loss in mmWave bands.
The directional antennas can be implemented by embedding many small antenna elements designed for mmWave in a limited physical space in mobile terminals as well as mmWave base stations (BSs).
However, the antenna directivity makes mmWave communication links vulnerable to link blockage caused by moving obstacles.
The link blockage suddenly penalizes the mmWave link budget by 20--30\,dB in the case of data transmission comprising the use of directional antennas\cite{haneda, flexible}.
The sudden and damaging degradation in the received power causes frequent interruptions within a transmission of streamed data, which is a crucial problem for VR/AR applications.

To overcome the blockage problem and provide reliable mmWave communications, a handover between multiple BSs is envisioned as a promising scheme\cite{multiAPs, reactive, umehira, polese2017improved, sun2018smart}.
By performing handovers at appropriate times, the decreased link budget can be compensated with another BSs.
In next-generation cellular networks, an increasing number of mmWave BSs will be deployed to ensure a line-of-sight (LOS) path between a mobile terminal and one of the deployed BSs; hence, designing a decision problem concerning when and to which BS a handover should be triggered, which is referred to as a \textit{handover decision problem}, is an important research direction.

In a handover decision problem, it is important to predict a future long-term performance, e.g., the time-average or cumulative sum of the data rates prior to performing a handover in order to avoid making myopic decisions\cite{mdp, mmwave_mdp, dhahri2014adaptive,otsuki_Q, icc2012_handoff, chang2008cross}.
This is because a handover involves a service disruption caused by procedures that are necessary for changing association and for data forwarding to a BS to which handover is performed\cite{liu2015smart}.
Performing handovers based on a short-term performance, i.e., making myopic decisions, results in frequent handovers that may cause the overall long-term performances to deviate\cite{otsuki_Q, dhahri2014adaptive}.
Thus, a future long-term performance in both the currently associated BS and the candidate BSs should be predicted prior to triggering a handover, and a handover decision rule should be formed such that the predicted performance is maximized.

In addition to the avoidance of redundant handovers, predicting future long-term performance is beneficial to avoid a lower data rate situation, particularly in mmWave communications, what is a main topic of this study.
Due to moving obstacles, mmWave links experience faster data rate variation compared to microwave links.
Given such constraints, the data rate provided by current BS may be lower than the rates provided by another BS before a handover execution is performed if the handover occurs after data rate variation.
As a result, severe loss of the data rate takes place.
By predicting future data rates within a longer time horizon, handover is performed to avoid data rate loss, so \textit{proactive handover} is beneficial rather than detrimental.

However, it is still challenging to predict the future long-term performance in mmWave links proactively under the condition that moving obstacles cause the rapid variation of received powers or data rates.
This is because the sudden variation exhibits little prior indications in the radio frequency (RF) signal domain such as received power samples and channel state information\footnote{With regard to the degradation of the received powers, there is a slight fluctuation in the received powers within 100\,ms prior to the degradation, which is known as diffraction effects\cite{double_knife, flexible, koda_measurement2, 11ad_channel}.
Hence, by analyzing the time-series of the received powers, we can predict the degradation from at most 100\,ms before the occurrence\cite{kaltiokallio2017three}.
Nonetheless, it is worthwhile utilizing the camera image domain for the two reasons.
First, as experimentally confirmed in this paper, based only on the variation, the degradation cannot be necessarily predicted in a proactive manner.
Second, the degradation should be predicted earlier because the service interruption incurred by a handover could be several hundreds of milliseconds long \cite{lin2017handoff}.}.
Thus, to predict the rapid variations in data rates or received powers proactively, we should utilize other information domains that provide more informative features for predicting such variations.


To address this challenge, this study develops a proactive framework wherein future data rate degradations caused by moving obstacles are predicted from several hundreds of milliseconds before the degradation occurs and the handover timings are optimized based on the predicted values. 
The key idea is to leverage the time consecutive camera images\footnote{
	We used depth images pixels of which are used to measure the distance between the obstacles and the camera \cite{depth}.
	Depth images allow us to obtain geometric relations between components within the scene.
	In the following discussion, we consider that the depth images are available to a network controller.
} and to use deep reinforcement learning (RL).
Time consecutive camera images comprise information of the spatiotemporal dynamics of moving obstacles, which exhibits informative features for predicting the future obstacle-caused degradation of data rates in mmWave links.
The optimization of the handover timings while predicting such future degradations based on camera images is a new challenge. 
We incorporate the usage of camera images into the RL-based handover frameworks (discussed in detail in the following section) by expanding the state space such that the state involves camera images.
 Moreover, by using a deep RL\cite{DQN}, we overcome the difficulty in handling the large dimensionality of the state space incurred by the state space expansion.

The most closely related work was submitted to the IEEE CCNC 2020 \cite{koda2019cooperative},
while the contributions of this paper are different from those in \cite{koda2019cooperative}.
As discussed later in detail, the main contribution of this study is the presentation of proactive prediction in handover decision problems by leveraging camera images.
Meanwhile, \cite{koda2019cooperative} addressed the issue of how to compensate for a blind spot of a single camera while applying the framework proposed in this paper and proposed a multi-camera operation.
Thus, the contribution of \cite{koda2019cooperative} is to demonstrate the feasibility of incorporating the multi-camera operation into the framework proposed in this paper.

The contributions of this paper are summarized as follows:
\begin{itemize}
	\item We highlight that the variation in the received powers before blockage events is not necessarily informative in predicting future data rate degradation in mmWave links.
	To confirm this, we obtained experimentally a received power time series that exhibits the variation and predicted the cumulative sum of future data rates with the RL method based on the state of received power obtained.
	\item Based on the following two ideas, we propose a proactive framework wherein handover timing is optimized while the degradation in data rate caused by obstacles is predicted within hundreds of milliseconds before degradation.
	The first idea is to expand the states such that the states comprise time-consecutive camera images, which provide informative features for predicting degradations, i.e., spatiotemporal dynamics of moving obstacles.
	The second idea is to leverage deep RL to overcome the computational complexity of learning the optimal handover policy incurred by the expanded state.
\end{itemize}

The rest of this paper is organized as follows.
Section~\ref{sec:rss_based} presents an experimental evaluation of the received-power-based prediction of the cumulative sum of the future data rates in a handover decision problem.
Section~\ref{sec:image_based} presents our image-based handover framework, which leverages time-consecutive camera images in a handover decision problem.
Finally, Section~\ref{sec:conc} presents concluding remarks.

It should be noted that Sections~\ref{sec:rss_based} and \ref{sec:image_based} are related to each other.
The former provides a baseline for the framework without camera images to be compared with the proposed image-based handover framework, and the latter details the image-based handover framework.
In concrete, in Section~\ref{sec:rss_based}, the problem of a received power-based handover framework summarized in the first contribution is highlighted.
This received power-based handover framework is referred to as baseline without camera images, and compared to the proposed image-based handover framework  in Section~\ref{sec:image_based}.
In Section~\ref{sec:image_based}, focusing on the highlighted problem, we propose the image-based handover framework presented in the second contribution.
Subsequently, we discuss the difference between the handover policies learned with and without camera images by comparing our image based-handover framework with the received power-based handover framework.

\section{Related Works}

\begin{table*}[!t]
	\centering
	\caption{Comparison of Handover-Related Previous Works}
	\label{table:related_works}
	\begin{tabular}{cccccccc}\toprule
		 & \cite{mdp, chang2008cross, sung2013predictive} &  \cite{otsuki_Q, dhahri2014adaptive, icc2012_handoff} & \cite{mmwave_mdp, zang2017mobility, zang2019managing} & \cite{koda_infocom} & \cite{proactive,proactive3,proactive4} & \cite{okamoto,okamoto_access} & \textbf{This paper}\\\midrule
		 Optimization of handover timings & Yes & Yes & Yes& Yes & No & No & \textbf{Yes}\\
		 Frequency band& Microwave & Microwave & mmWave & mmWave & mmWave & mmWave & \textbf{mmWave}\\
		 Usage of camera images&No & No & No & No & Yes & Yes & \textbf{Yes}\\
		 Proactive blockage prediction & No & No & Yes & Yes & Yes & Yes & \textbf{Yes} \\
		 Approach & DP & RL  & DP & RL  & Heuristic & SL & \textbf{RL}\\\bottomrule
	\end{tabular}
\end{table*}

\subsection{Handover Decision Problems}
\label{subsec:handover-decision-problems}

In many studies,  handover decision-making problems or cell selection problems in heterogeneous microwave networks or millimeter wave networks were formulated with the objective of maximizing the future long-term performance \cite{mdp, mmwave_mdp, dhahri2014adaptive,otsuki_Q, icc2012_handoff, chang2008cross, sung2013predictive}.
The authors of \cite{mdp, chang2008cross, sung2013predictive} designed the optimal cell selection problem in heterogeneous wireless networks with the objective of maximizing the weighted sum of the network bandwidth and network delay via the Markov decision process (MDP) models or optimal control models.
The optimal strategies are provided via dynamic programming (DP) techniques.
In \cite{mmwave_mdp}, optimal cell selection in mmWave networks was proposed to maximize the long-term throughput or total received data in a mobile terminal using a similar approach.
The authors of \cite{icc2012_handoff,otsuki_Q, tan2014cell, dhahri2014adaptive} applied an RL algorithm to learn the optimal cell selection with the objective of maximizing the long-term quality of experiences or channel capacities, wherein an optimal strategy of cell selection can be learned without prior knowledge of the transition probability of the channel states or received powers.
However, in the aforementioned studies, a decision process was considered wherein a decision maker makes a decision based on a current network state such as the channel information, received power, or network bandwidth.
These studies did not detail the challenge of predicting the future long-term performance in mmWave links under the condition of moving obstacles causing blockage effects and received powers at a station (STA) or the BSs and the data rates in the mmWave links undergoing rapid degradation.

Other works have addressed handover decision-making problems in mmWave networks by using user mobility information or pedestrian mobility information\cite{zang2017mobility, zang2019managing, koda_infocom}.
User mobility information facilitates the prediction of future data rates in mmWave links with blockage effects that occur when users are entering areas blocked by static obstacles\cite{zang2017mobility, zang2019managing}.
However, the proactive prediction of the data rate degradations caused by moving obstacles is not addressed.
In our previous work\cite{koda_infocom}, we addressed handover decision problems based on the positions and velocities of a moving pedestrian.
However, the proposal is not applicable to handover decision problems wherein more pedestrians cause blockage effects because of the challenge of capturing the spatial features of each pedestrian such as their height or shape.
In contrast, our current proposal uses camera images that comprise spatial information, thereby capturing the spatial features of moving obstacles.

\subsection{Camera Image-Based Frameworks in mmWave Networks}
The authors of \cite{proactive,proactive3,proactive4} have conceptualized a camera-assisted proactive handover system for mmWave networks.
The camera images are employed to predict the occurrence of blockage effects caused by pedestrians approaching a LOS path between a BS and an STA.
The experiments conducted in these works demonstrated that using camera images, a handover can be triggered several seconds before blockage the occurrence of the blockage effects.
However, the methods embedded in the experiments are focused on predicting the timings at which blockage effects occur, and they do not quantitatively predict the future data rate degradation caused by pedestrians. 
As discussed in the previous section, the optimal handover requires a prediction of the future long-term performances; hence, the aforementioned methods cannot provide the optimal solution to handover decision problems. 

Motivated by the issue, a novel method for quantitatively predicting a future received power value in mmWave communications was proposed in\cite{okamoto_access}.
The method predicts a received power value from several hundreds of milliseconds before the value is observed.
In this method, camera images are mapped, via a supervised learning (SL) technique, to a future received power value that is obtained several hundreds of milliseconds after the camera images are obtained.
However, the prediction method in \cite{okamoto_access} is not specific to handover decision problems.
While the method in \cite{okamoto_access} can be used to predict a future data rate at a certain time period, the optimization of the handover timings requires a different prediction, i.e., the prediction of the expected cumulative sum of future long-term data rates as confirmed in the previous studies discussed in Section~\ref{subsec:handover-decision-problems}.
Thus, the method in \cite{okamoto_access} cannot be necessarily adopted directly in  handover decision problems.
Table~\ref{table:related_works} summarizes the main aspects of the previous works related to this paper.

\section{Received Power-Based Handover Framework}
\label{sec:rss_based}

The main objective of this section is to highlight that the future degradation of data rates in mmWave links caused by moving obstacles cannot necessarily be predicted based only on a variation in received powers.
To illustrate this point, we perform a prediction of the cumulative sum of the future data rates using RL with the state information of the experimentally obtained received powers.
We will refer to the received power-based handover framework as a baseline without camera images, to be compared with the proposed image-based handover framework in Section~\ref{subsec:results}.
First, we provide an overview of the RL.
Then, we present the decision process considered in this experiment.
Finally, we provide an experimental study of the prediction based on the received powers.

\subsection{Overview of RL}

General RL algorithms are performed over an MDP.
An MDP consists of the following four elements:
a state space $\mathcal{S}$, an action space $\mathcal{A}$, a reward function $r:\mathcal{S}\times\mathcal{A}\times\mathcal{S}\to\mathbb{R}$, and transition probabilities $q:\mathcal{S}\times\mathcal{A}\to\Omega(\mathcal{S})$, where $\Omega(\mathcal{S})$ denotes the collection of the probability distribution over $\mathcal{S}$.
At each decision epoch $t\in\mathbb{N}$, a decision maker observes the state information $s_t\in\mathcal{S}$.
Subsequently, the decision maker selects an action on the basis of the {\it policy} $\pi:\mathcal{S}\to\mathcal{A}(s_t)$, where $A(s_t)\subseteq \mathcal{A}$ denotes the set of possible actions when the state $s_t$ is observed.
Given the current state $s_t$ and selected action $a_t\in\mathcal{A}(s_t)$, the state transitions to $s_{t + 1}\in\mathcal{S}$ at the next decision epoch $t + 1$ according to the transition probability $q(s_{t + 1}, s_t, a_t)$; thereafter, the decision maker is given a reward $r(s_{t + 1}, a_{t}, s_t)$.

The objective of the decision maker is to determine the optimal policy $\pi^{\star}$ that maximizes the total expected discounted reward. 
The optimal policy satisfies the following condition:
\begin{multline}
	\label{eq:optimal_policy}
	\mathbb{E}\!\left[\,\sum_{t' = 0}^{\infty}\gamma^{t'}r\bigl(s_{t + t' + 1}, \pi^{\star}(s_{t + t'}), s_{t + t'}\bigr)\,\middle|\,s_t = s\right]
	\\
	\geq \mathbb{E}\!\left[\,\sum_{t' = 0}^{\infty}\gamma^{t'}r\bigl(s_{t + t' + 1}, \pi(s_{t + t'}), s_{t + t'}\bigr)\,\middle|\,s_t = s\right],
\end{multline}
$\forall s\in S$ and $\forall\pi$, where
$\gamma\in[0, 1)$ represents the discount factor.
In the MDP wherein $\mathcal{S}$ and $\mathcal{A}$ are both countable non-empty sets, there exists at least an optimal policy\cite{sutton}.

To obtain the optimal policy in an MDP, it is sufficient to obtain the optimal action-value function $Q^{\star}:\mathcal{S}\times\mathcal{A}\to\mathbb{R}$.
The optimal action-value function is defined as follows:
\begin{multline}
	\label{eq:opt_q_func}
	Q^{\star}(s, a) \coloneqq \mathbb{E}_{s'}\!\left[r(s', a, s) + \gamma V^{\star}(s')\,\vert\,s, a\right],\\ s\in \mathcal{S}, a\in\mathcal{A}(s),
\end{multline}
where $\mathbb{E}_{s'}[\,\cdot\,\vert\, s, a\,]$ denotes the expectation operator under the transition probability $q(s', s, a)$ and $V^{\star}(s)$ denotes the left-hand side in \eqref{eq:optimal_policy}.
This is attributed to the fact that the optimal action-value function is related to the optimal policy as follows\cite{sutton}:
\begin{align}
	\label{eq:greedy_optimal}
	\pi^{\star}(s) = \argmax_{a\in\mathcal{A}(s)}Q^{\star}(s, a).
\end{align}
In other words, the policy that selects the action that maximizes $Q^{\star}(s, a)$ is optimal.
In this study, the optimal action-value function is learned using deep RL\cite{DQN}.

\subsection{States, Actions, Rewards, and State Transition Rules}
We present the decision process considered in this experiment by detailing the states, actions, rewards, and state transition rules.
In the process, a network controller makes handover decisions in the mmWave networks based on the received power values. 
We consider a mmWave network wherein multiple mmWave BSs and an STA are deployed.
There exist obstacles that block the LOS path between the STA and the BS associated with the STA.
We also consider the decision of whether a handover should be triggered with respect to the time length of service disruption.
The communication between the BS and STA can be disrupted because of the necessary procedures for the association, which involves beam alignment and for data forwarding to a BS to which a handover is performed\cite{handover_cost, lin2017handoff, hassanieh2018fast}.
We define the duration for which the communication is disrupted as the service disruption time $T_{\mathrm{dis}}$.

It should be noted that in many existing studies\cite{mdp, chang2008cross, mmwave_mdp, zang2017mobility, zang2019managing}, the handover decision process was formulated as an MDP, although it was assumed that the interval between the decision epochs was several seconds long, which is longer than a realistic service disruption time of several tens or hundreds of milliseconds\cite{handover_cost}.
Hence, the service disruption occurs within an interval between the successive decision epochs.
However, the assumption of the large interval is not suitable for predicting the blockage effects that moving obstacles cause within several hundred milliseconds\cite{11ad_channel}.
Hence, we reformulate the problem wherein an interval between the successive decision epochs is shorter than several tens or hundreds of milliseconds, and several decision epochs could be within a service disruption.

\subsubsection{States}
	For the network controller to detect blockage effects based on received powers, we design the states such that they include the received power values.
	Let the number of time-consecutive received power values used in making handover decisions be denoted by $N$.
	We set the state space as follows:
		  \begin{align}
			\label{eq:state_set_rss}
		      \mathcal{S}_{\mathrm{rp}}\coloneqq \underbrace{\mathcal{P}\times\dots\times\mathcal{P}}_{N}\times\mathcal{J}\times\mathcal{C}.
		  \end{align}
	In \eqref{eq:state_set_rss}, $\mathcal{P}\subseteq \mathbb{R}^{J}$ denotes the set of all possible received powers observed at all BSs, $\mathcal{J}\coloneqq\{1,\dots,J\}$ denotes the set of the BS indices, and $ \mathcal{C} \coloneqq \{\,c\mid c\in\mathbb{Z}, 0\leq c\leq \lfloor T_{\mathrm{dis}}/\tau \rfloor\,\}$ denotes the set of the remaining decision epochs until the service disruption time ends, where $J$ denotes the number of the deployed BSs, $\lfloor \cdot \rfloor:\mathbb{R}\to\mathbb{R}$ denotes the floor function, and $\tau$ denotes the interval between the successive decision epochs.
	
	Let $s_t = (p_{t}, p_{t - 1},\dots, p_{t - N + 1}, j_t, c_t)\in\mathcal{S}_{\mathrm{rp}}$ denote the state at the decision epoch $t$.
	The element $p_{t - k}\in\mathcal{P}$ for $k\in \{0, 1, \dots, N - 1\}$ is set as the received power observed at the decision epoch $t - k$.
	The element $j_t\in\mathcal{J}$ is set as the index of the BS associated with the STA.	  
	The element $c_t\in \mathcal{C}$ is set as the number of remaining decision epochs that the network controller experiences until the service interruption ends.
	When the decision epoch is not within the service disruption time, $c_t$ is set as zero.
	
	\subsubsection{Actions} 
	      We let the set of possible actions $\mathcal{A}(s_t)$ be as follows:
				\begin{align}
					\label{eq:action_set}
		      \mathcal{A}(s_t)\coloneqq
		      \begin{cases}
			      \mathcal{J}, & c_t = 0;    \\
			      \{j_t\},       & c_t \neq 0.
		      \end{cases}
	      \end{align}
	      In other words, the controller selects one of the BSs when the decision epoch is not within the service disruption time; otherwise, the controller selects only the index of the BS to which a handover is performed.

	\subsubsection{Reward}
	 We set the reward as a performance metric in the link provided by the BS that is currently associated with the STA with the exception that when the next decision epoch $t + 1$ is within the service disruption duration, we set the reward as zero as follows:
	      \begin{align}
		      \label{eq:reward}
		      r(s_{t + 1}, a_t, s_t) \coloneqq
		      \begin{cases}
			      R_{j_{t + 1}, t + 1}, & c_{t + 1} = 0;    \\
			      0,      & c_{t + 1} \neq 0.
		      \end{cases}
	      \end{align}
		  In \eqref{eq:reward}, $R_{j_{t + 1}, t + 1}$ denotes the performance metric in the link provided by BS $j_{t + 1}$ at $t + 1$.
			In the performance evaluation, we set $R_{j_{t + 1}}$ as the achievable data rate provided by BS $j_{t + 1}$ as discussed in Section~\ref{fig:subsec_rss_based_experiment}.
			
	\subsubsection{State Transition}
	The state transition to the next state is as follows.
	Let the state at epoch $t + 1$ be $s_{t + 1} = (p_{t + 1}, p_t, \dots, p_{t - N + 2}, j_{t + 1}, c_{t + 1})\in\mathcal{S}_{\mathrm{rp}}$.
	Evidently, the received power values $(p_{t + 1}, p_{t}, \dots, p_{t - N + 2})$ at $t + 1$ are updated by concatenating the received power values at $p_{t + 1}$ with the current values $(p_{t}, p_{t - 1}, \dots, p_{t - N + 1})$ and removing the oldest value $p_{t - N + 1}$.
	Based on the definition of the state, the term $j_{t + 1}$ is determined as follows:
	\begin{align}
		\label{eq:j_t_transition}
		j_{t + 1} = a_t.
	\end{align}
	The term $c_{t + 1}$ is determined as follows:
	\begin{align}
		\label{eq:c_t_transition}
		c_{t + 1} =
		\begin{cases}
			c_t - 1;                          & \text{$c_t \neq 0$},       \\
			\lfloor T_{\rm dis}/\tau \rfloor; & \text{$c_t = 0, a_t\neq j_t$}, \\
			0;                              & \text{$c_t = 0, a_t = j_t$}.
		\end{cases}
	\end{align}

	It should be noted that without knowing the transition probabilities, we learn the optimal action-value function using deep RL\cite{DQN}.
	To learn the optimal policy, we only require transition samples $(s_t, a_t, r_t, s_{t + 1})$ that can be obtained while making decisions in the learning procedure.		

We detail an example of the temporal transition of the decision process.
We consider that at the decision epoch $t$, $s_{t} = (p_{t}, p_{t - 1}, \dots, p_{t - N + 1}, 1, 0)$, i.e., the received power values $(p_{t}, p_{t - 1}, \dots, p_{t - N + 1})$ are available, the STA is associated with BS~1, and the decision epoch is not within the service disruption time.
If the controller selects action $a_t\neq 1$, i.e., a handover is performed, then the state transitions to $s_{t + 1} = (p_{t + 1}, p_{t}, \dots, p_{t - N + 2}, a_{t}, \lfloor T_{\rm dis}/\tau \rfloor)$.
The controller is subsequently given a reward of zero because $c_t  = \lfloor T_{\rm dis}/\tau \rfloor \neq 0$ (see \eqref{eq:reward}).
In this case, until the service disruption time ends, the controller selects action $a_{t}$, is given a reward of zero, and decreases the last element of the state by one.
Conversely, if the controller selects action $a_t = 1$, i.e., the handover is not performed, then the state transitions to the state $s_{t + 1} = (p_{t + 1}, p_{t}, \dots, p_{t - N + 2}, 1, 0)$ and the controller is then given the reward $R_{1, t + 1}$.

\subsection{Experimental Evaluation}
\label{fig:subsec_rss_based_experiment}

\begin{figure}[!t]
	\centering
	\includegraphics[width = 0.7\columnwidth]{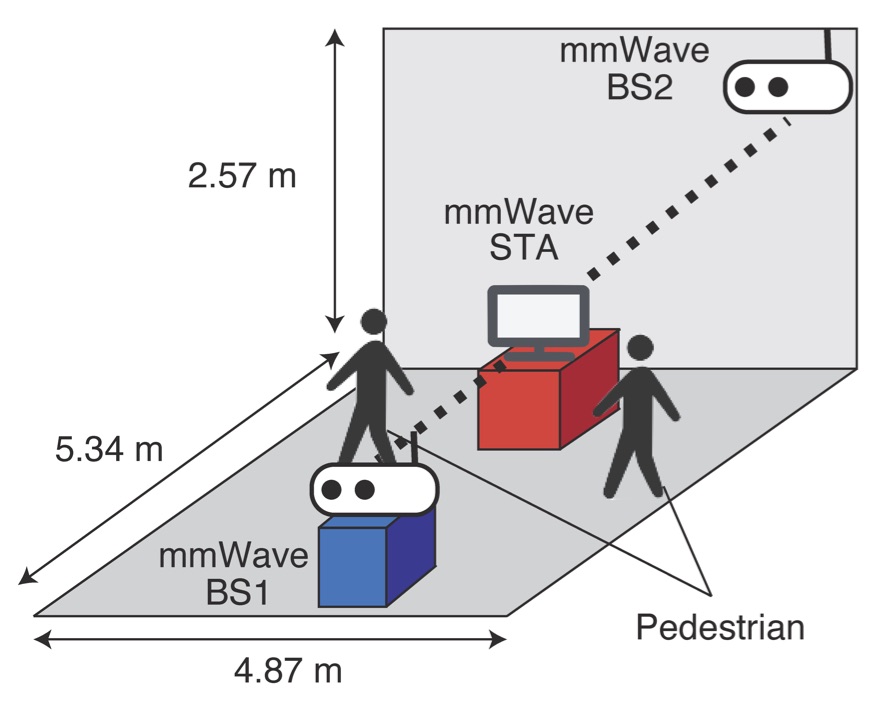}
	\caption{Experimented scenario of mmWave links.}
	\label{fig:simulated}
\end{figure}

\begin{figure}[!t]
	\centering
	\subfigure{\includegraphics[width = 0.58\columnwidth]{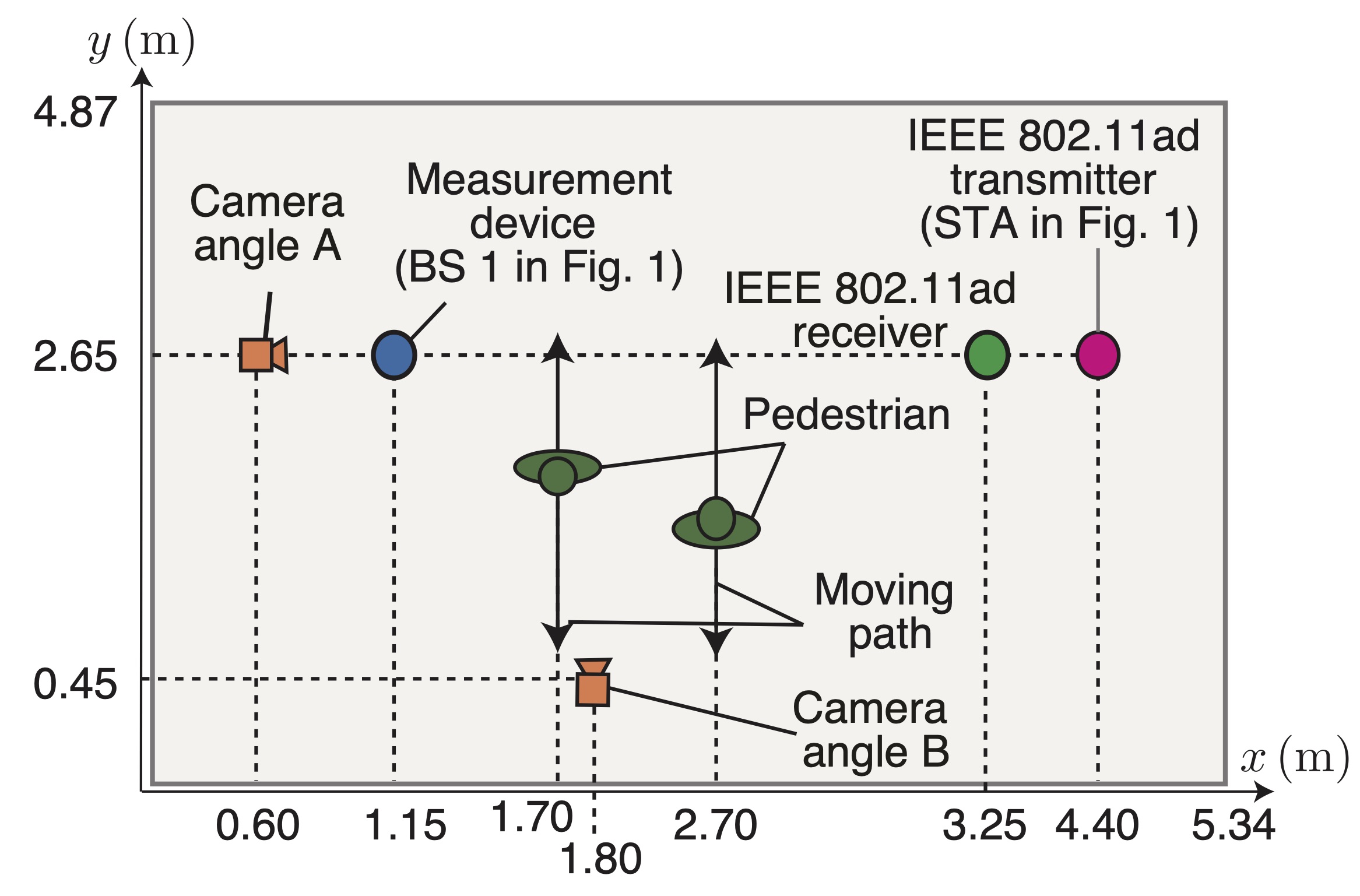}}\
	\subfigure{\includegraphics[width = 0.4\columnwidth]{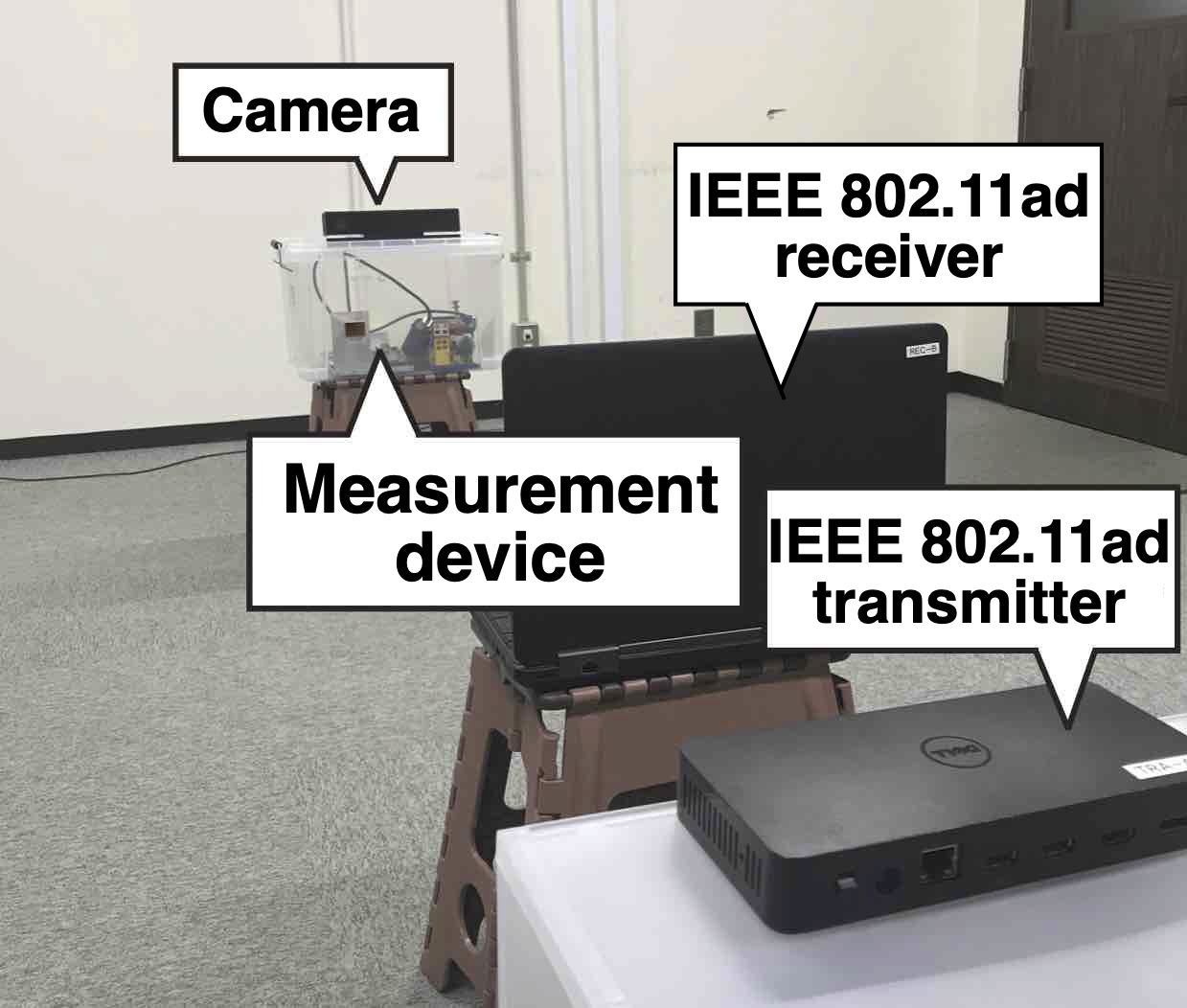}}
	\caption{Top view of the measurement environment (left) and measurement setup showing the mmWave transmitter, measurement device, and camera placed at the position A (right).
	The measurement device and mmWave transmitter correspond to BS~1 and the STA in Fig.~\ref{fig:simulated}, respectively.}
	\label{fig:mes}
\end{figure}
\begin{figure}[t]
	\centering
	\includegraphics[width=0.7\columnwidth]{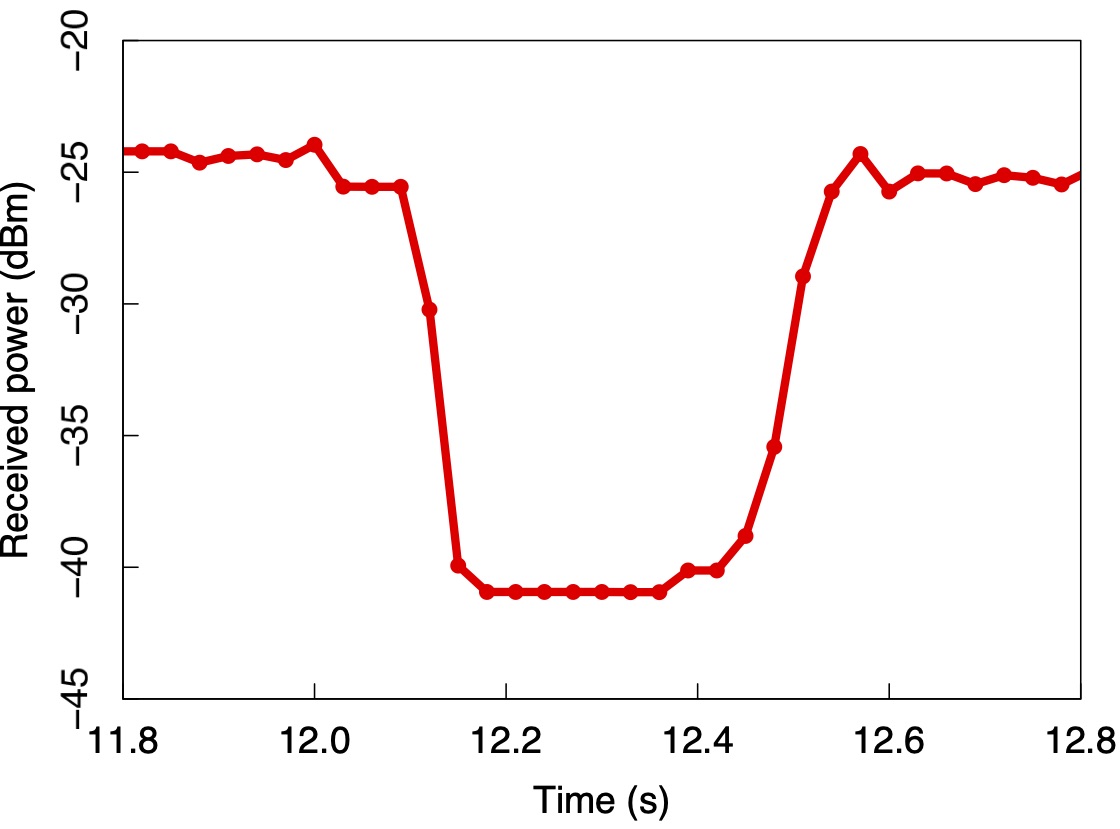}
	\caption{An example of the variation of the received power in a blockage effect.}
	\label{fig:channel}
\end{figure}

\subsubsection{Evaluated Scenario}
\label{secV:subsec:evaluated_scenario}

As shown in Fig.~\ref{fig:simulated}, two BSs and an STA are deployed in an indoor room whose length, width, and height are 4.87\,m, 5.34\,m, and 2.57\,m, respectively.
The size of the room corresponds to that of the room where the measurement in the next section was conducted.
The two BSs are operated over the 60\,GHz channel.
The STA is initially associated with the BS that observes a higher received power as compared to that of the counterpart when there are no obstacles within the deployed area.
We term the BS that is initially associated with the STA as BS~1 and the other as BS~2.
BS~2 is a candidate BS for a case in which the link between BS~1 and the STA is blocked by obstacles.
In the following discussion, we detail the considered scenario including coverage area, STA mobility, channel characteristics, beamforming, initial access procedure, and beam tracking.

\vspace{0.2em}\noindent\textbf{Coverage area:} A BS covers an entire room in Fig.~\ref{fig:simulated} at least in a LOS condition, which is examined as follows.
We determine whether a BS can cover a certain position or not by examining whether a commercially available IEEE 802.11ad equipped transmitter and receiver can associate with each other.
In this examination, we validated that the transmitter can associate with the receiver placed 10\,m apart from the transmitter in a LOS condition.
Because the maximum distance between two positions is 7.67\,m in the room, we can say that a BS covers the room in Fig.~\ref{fig:simulated} at least in a LOS condition, which is sufficient for performing the evaluation.

\vspace{0.2em}\noindent\textbf{STA mobility:} In this evaluation, the received power varies only because of moving obstacles, not because of the STA movements to focus on the sudden variation of the data rate caused by moving obstacles.
To obtain such received power samples, the experiment is arranged such that the position of the STA is static, and a handover is performed to compensate the degraded data rate caused by the obstacles rather than to support the STA mobility.
The evaluation in such a scenario is sufficient for the two objectives of this study: highlights that the such sudden variations of the data rate cannot be predicted from a received power time series (provided in Section~III) and demonstrates the feasibility of the proactive prediction on such variations achieved by camera images (provided in Section~IV).
This scenario of the static STA is reasonable for certain realistic application such as the transmission of streaming data to a wireless monitor in an office, discussed in Section~IV-B1 in detail.

\vspace{0.2em}\noindent\textbf{Channel characteristics between BS~1 and STA:} The channel between BS~1 and STA is based on the measurement in the next section, where the overall characteristics are similar to ``dynamic 60\,GHz radio channel''\cite{11ad_channel} in terms of the variation of the received powers only due to the moving obstacles.
An example of the variation of the received power in a blockage effect is shown in Fig.~3, and the overall characteristics are as follows.
In a blockage effect, the received power decreases by approximately 15\,dB within 50--200\,ms
Subsequently, the received power remains constant for 200--300\,ms, and then it recovers to the original value within 50--200\,ms.

\vspace{0.2em}\noindent\textbf{Blockage distributions:} Because the channel between BS~1 and STA is based on the measurement provided in the next section, the distributions characterizing the blockage events also depend on the measurement.
To quantitatively characterize the blockage events in the measurement, in Table~\ref{table:blockage_parameters}, we provide the estimated distribution parameters of the following five variables, which are essential in characterizing blockage events\cite{11ad_channel,peng2018statistical}.
In Table~\ref{table:blockage_parameters}, $t_{\mathrm{decay, 5\,dB}}$ denotes the duration in which signal attenuation level increases from 0 dB to 5 dB, $t_{\mathrm{rise, 5\,dB}}$ denotes the duration in which signal attenuation level decreases from 5 dB to 0 dB, $A_{\mathrm{mean}}$ is the mean signal attenuation, and $t_{\mathrm{D}}$ denotes the duration of the blockage event.
The value $t_{\mathrm{LOS}}$ is the duration wherein a LOS condition sustained.
The definitions of the values follow the works of \cite{11ad_channel} and \cite{peng2018statistical}.
The choice of the distribution functions is based on \cite{11ad_channel} or \cite{peng2018statistical}, and the parameters of the distributions are determined by the maximum-likelihood estimation.
The blockage events occurred within the 21\% time-length relative to the whole measurement time.

\begin{table}[!t]
	\caption{Estimated distribution parameters characterizing blockage events observed in measurement}
	\label{table:blockage_parameters}
	\begin{center}
		\begin{tabular}{ccc}\toprule
			Values & Distribution & Estimated Distribution Parameters\\\midrule
			$t_{\mathrm{decay, 5\,dB}}$ &Gaussian\cite{11ad} & mean: $0.059$, \\
			& & standard deviation: $0.034$\\
			$t_{\mathrm{rise, 5\,dB}}$ &Log Normal\cite{11ad} & log mean: $-3.01$, \\
			&&log standard deviation: $0.195$\\
			$A_{\mathrm{mean}}$ &Gaussian\cite{11ad} & mean: $14.2$, \\
			&&standard deviation: $2.08$\\
			$t_{\mathrm{D}}$ &Weibull\cite{11ad} & scale: $0.553$, shape: $4.08$\\
			$t_{\mathrm{LOS}}$ &Weibull\cite{peng2018statistical} & scale: $2.31$, shape: $1.51$\\
			\bottomrule
		\end{tabular}
	\end{center}
\end{table}

\vspace{0.2em}\noindent\noindent\textbf{Channel characteristics between BS~2 and STA:} Meanwhile, the mmWave channel between BS~2 and STA is static, and it is assumed that BS~2 is free of blockages.
The assumption is reasonable given that a network controller is likely to perform a handover to a BS that is not blocked by obstacles.
In the following discussion, it is considered that BS~2 is at a position where pedestrians cannot block the path between the STA and BS~2, and the received power at BS~2 is constant over time.
Since the focus of the evaluation is on a blockage effects between STA and BS~1, we detail the link between the STA and BS~1 in the following discussion.

\vspace{0.2em}\noindent\textbf{Beamforming:}
The STA and BS~1 communicate with each other with directional antennas.
Because channel characteristics between BS~1 and STA are based on the measurement, the antenna gain is also attributed to the measurement equipment.
In the next section, we detail antenna gains of a transmitter and measurement device, which corresponds to the STA and BS~1, respectively.

\vspace{0.2em}\noindent\textbf{Initial access procedure:}
Prior to the evaluation, we established the beam of STA based on an initial access procedure termed as iterative beam search method\cite{hassanieh2018fast}, which is used in the IEEE 802.11ad standard.
This initial access procedure is because we used the a commercially available IEEE 802.11ad equipped transmitter as the STA in the measurement.
Meanwhile, we established the beam of BS~1 manually such that the beams of BS~1 and the STA point towards each other.
We discuss the procedure of establishing the beam directions of the STA and BS~1 in detail in the next section.

\vspace{0.2em}\noindent\textbf{Beam tracking:}  
	In the next section, we conduct the measurement such that the STA and BS~1 do not perform a beam tracking.
	The aim is to eliminate the variation of the received powers due to beam tracking whose mechanisms depend on manufacturers and thereby to focus only on the sudden variations of the received powers due to moving obstacles.
	In the next section, we detail how the measurement is conducted such that the beam tracking is not performed.

\subsubsection{Measurement Setup}
\label{secV:subsec:experiment_setup}
We set up an IEEE 802.11ad equipped transmitter/receiver, a measurement device, and a camera as shown in Fig.~\ref{fig:mes}.
The transmitter and a measurement device correspond to the STA and BS~1, respectively.
The transmitter and measurement device is place at the height of 0.70 and 0.85\,m, respectively.
The camera is placed at either position $(0.60, 2.65)$ and $(1.80, 0.45)$ and at the heights of 1.50\,m and 1.25\,m, respectively to obtain the a dataset with two different camera angles.
The angle from the former position is termed angle~A while the latter is termed angle~B.
The measurement device is equipped with a horn antenna with directivity gain of 24\,dBi and the half-power beam width (HPBW) of 11\,degree while the transmitter is equipped with an array antenna with size of 16, directivity gain of approximately 8\,dBi, and HPBW of approximately 15\,degree\cite{d5000}.

The beam directions of the measurement device and transmitter are established from the following procedure to configure the beams of the measurement device and transmitter such that the beams point towards each other.
First, the transmitter and receiver in Fig.~\ref{fig:mes} performed, in a LOS condition, an iterative beam search wherein the beam pair was searched with a two-stage beam scanning\cite{hassanieh2018fast} such that the receiver can benefit from the maximum received power.
Through the procedure, the beam of the transmitter is configured such that the beam points towards the receiver.
Subsequently, we placed the measurement device behind the receiver such that the horn antenna attached with the measurement device faced for the transmitter.
Since the beam of the transmitter also points towards the measurement device, the beams of the transmitter and measurement device point towards each other.

We conduct the measurements as in \cite{koda_measurement2} and obtain the received powers and camera images.
The mmWave transmitter transmits signals at the carrier frequency of 60.48\,GHz to the receiver, and subsequently, the measurement device behind the receiver measures the power of a part of the signals\cite{koda_measurement2}. 
The transmitted signals are considered as uplink signals from the STA to BS~1.
In this environment, two pedestrians walk along the moving path in Fig.~\ref{fig:mes} and obstruct the path between the transmitter and measurement device.
Tables \ref{table:impl} summarizes the experimental equipment and parameters associated with the experiment.

We conduct the measurement such that the beam tracking between the measurement device and transmitter is not performed.
This eliminates the variation of the received powers due to beam tracking whose mechanisms depend on manufacturers, and we can thereby focus only on the sudden variations of the received powers due to moving obstacles.
The details are as follows:
The measurement device was located behind the receiver, and the pedestrians traveled between the receiver and measurement device indicated in Fig.~\ref{fig:mes}.
This arrangement prevented the receiver and transmitter from performing beam tracking because the received power at the receiver was not altered even when the LOS path between the receiver and measurement device was blocked.
In this situation, the beam direction of the transmitter is almost fixed.
Consequently, the beam directions in the transmitter and measurement device were also fixed, wherein the beam tracking between them was not performed.

It should be noted that the camera images obtained in this experiment are not used in this evaluation but are used in the next section.
This evaluation provides the baseline that does not utilize camera images and only uses the time series of received powers to decide handover timings.

\begin{table}[!t]
	\caption{Experimental Equipment and Parameters}
	\label{table:impl}
	\begin{center}
		\begin{tabular}{cc}	\toprule
			IEEE 802.11ad transmitter & Dell Wireless Dock D5000                  \\
			IEEE 802.11ad receiver & Dell Latitude E5540                  \\
			Spectrum analyzer  & Tektronix RSA306                          \\
			Down-converter     & Sivers IMA FC2221V                        \\
			Antenna            & Sivers IMA Horn antenna, 24\,dBi          \\
			Depth camera       & Microsoft Kinect  \\
			& for Windows (Model:1656)\\
			Channel                    & 60.48\,GHz           \\
			Sampling frequency         & 56\,MHz              \\
			Transmit antenna gain      & 10\,dBi \cite{d5000} \\
			Receive antenna gain       & 24\,dBi              \\
			Measurement bandwidth $W$ & 40\,MHz, 20\,MHz    \\
			\bottomrule
		\end{tabular}
	\end{center}
\end{table}

\begin{table}[t]
	\caption{Parameters Associated with RL}
	\label{table:param}
	\centering
		\begin{tabular}{cc}	\toprule
			Discount factor, $\gamma$                                                           & 0.99                                    \\
			
			Number of obtained received powers $T$                                                  & 16860                                   \\
			Number of received powers used for learning $T'$                                                  & 13500  \\
			Number of iterations
			 of learning and evaluation                                                              & 1000                                    \\
			Exploration rate $\epsilon$                                                         & 1--0.01 \\
			&(Decreased by 0.01 \\
			& per iteration)\\
			Number of received power values in state $N$                        & 2                                      \\
			Interval between successive decision epochs $\tau$                                  & $30$\,ms                       \\
			Noise power spectral density $\sigma^2$ & $-$173\,dBm/Hz\\ 
			Received power at BS~2, $p_{2, t}$                                    & $-$129\,dBm (const.)                    \\
			Batch size\cite{DQN}                                                                    & 32                                      \\
			Frequency of updating the target network\cite{DQN}                                          & 10000                                 \\
			\bottomrule
		\end{tabular}

\end{table}

\subsubsection{Procedure of Performing Decision Process}
\label{secII:subsubsec:simulate_MDP}
We divide the received powers into two parts, and the individual parts are used for the learning and performance evaluation, respectively.
Let the obtained received power values be denoted by $\bigl(p_{1, t}\bigr)_{t\in\mathcal{T}}$, where $p_{1, t}$ denotes the received power obtained at the $t$th observation, and $\mathcal{T} = \{1, 2, \dots, T\}$ denotes the set of the time indices.
We divide $\mathcal{T}$ into the following two subsets: $\mathcal{T}_1 = \{1, 2, \dots, T'\}$ and $\mathcal{T}_2 = \{T' + 1, T' + 2, \dots, T\}$, where $1 < T' < T$.
We use $(p_{1, t})_{t\in\mathcal{T}_1}$ to learn the optimal action-value function and $(p_{1, t})_{t\in\mathcal{T}_2}$ to evaluate the learned policy.
In the following discussion, we denote $p_t$ as the received power values observed at BS~1 and at BS~2, i.e., $(p_{1, t}, p_{2, t})$, where $p_{2, t}$ is the received power value observed at BS~2 and is constant $\forall t \in \mathcal{T}$.

We simulate the decision process in the learning procedure using $(p_{t})_{t\in\mathcal{T}_1}$.
The decision epoch is set as the time step at which a received power value is obtained.
The decision process starts at the time step at which $p_{1, N}$ is observed.
The STA is initially associated with BS~1 and the time at which the process starts is not within a service disruption time, i.e., $j_1 = 1$  and $c_1 = 0$.
Thus, the state $s_N$ is set as $(p_{N}, p_{N - 1}, \dots, p_{1}, 1, 0)$.
The action $a_N$ is selected according to a heuristic $\epsilon$-greedy policy\cite{DQN}; the next state $s_{N + 1}$ is then set such that it includes the images $(p_{N + 1}, p_N, \dots, p_{2})$, $j_{N + 1}$, and $c_
{N + 1}$, where $j_{N + 1}$ and $c_{N + 1}$ are determined based on $a_N$ as shown in \eqref{eq:j_t_transition} and \eqref{eq:c_t_transition}.
The procedure is iterated, and it then ends when the state includes the received power values $p_{T' - 1}$.

The performance metric $R_{j, t + 1}$ for $j\in\mathcal{J}$ in \eqref{eq:reward} is set as the achievable data rate provided by BS~$j$, which is associated with the STA and is calculated as follows.
The metric $R_{j, t + 1}$ is calculated by the Shannon capacity formula with the received power value $p_{j, t + 1}$ as follows:
\begin{align*}
	R_{j, t + 1} = W\log_{2}\!\left(1 + \frac{p_{j, t + 1}}{\sigma^2 W}\right),
\end{align*}
where $\sigma^2$ denotes the noise power spectral density. 
It should be noted that the metric at  $R_{2, t + 1}$ is set as a constant value based on the assumption that the received power at BS~2 is a constant over time.

We evaluate the performance of the learned policy in the following step termed as performance test.
In the performance test, we simulate a decision process using the same procedure as the learning procedure with the exception that we use $(p_{t})_{t\in\mathcal{T}_2}$, and the action is selected according to a greedy policy\cite{sutton}.
We calculate the time average of the reward as a performance metric of the learned policy.
We then iterate the learning and evaluation using the same data set.
We evaluate the policy that achieves the highest average reward throughout the iterations.

It should be noted that the handover policy is learned via deep RL\cite{DQN} with a neural network (NN) that is different from that shown in Fig.~\ref{fig:nn_arc} (discussed later).
We simplify the NN architecture because the input of the NN in this scenario comprises several elements---the four elements in the evaluation.
We replace the combination of the convolutional neural network (CNN) and long short-term memory (LSTM) in Fig.~\ref{fig:nn_arc} with a fully connected multi-layer with eight hidden units and 32 output units, where the two layers are activated using rectified linear units\cite{goodfellow}.
The parameters associated with the deep RL are summarized in Table~\ref{table:param}.

\begin{figure}[t]
	\centering
	\subfigure[Time series of achievable data rate provided by BS~1.]{\includegraphics[width = \columnwidth]{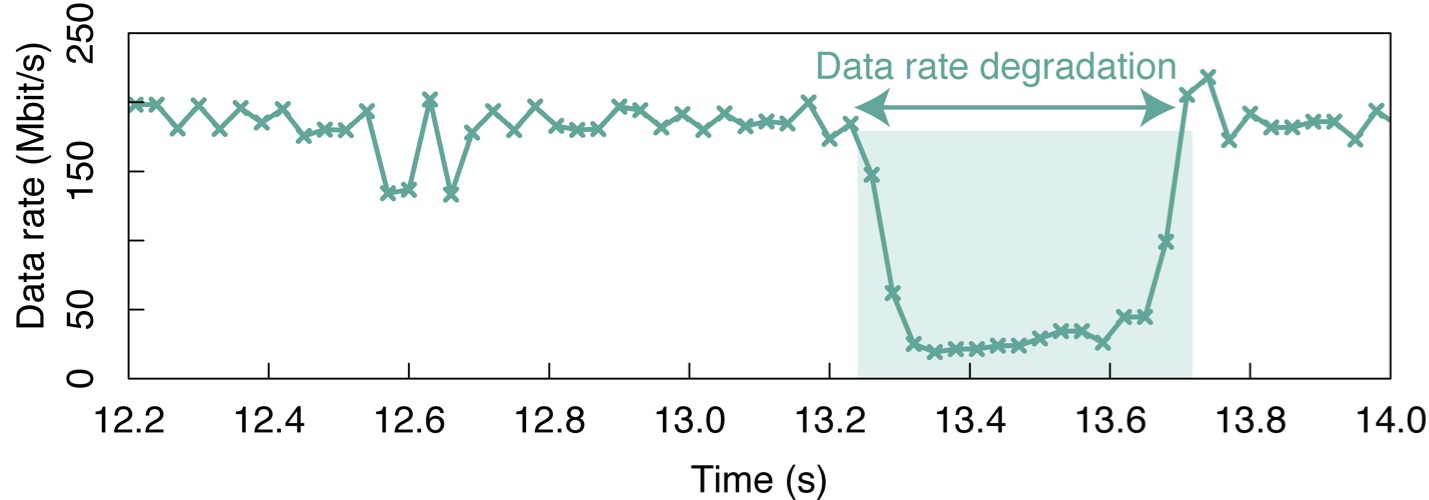}}
	\subfigure[Time series of learned action values.]{\includegraphics[width = \columnwidth]{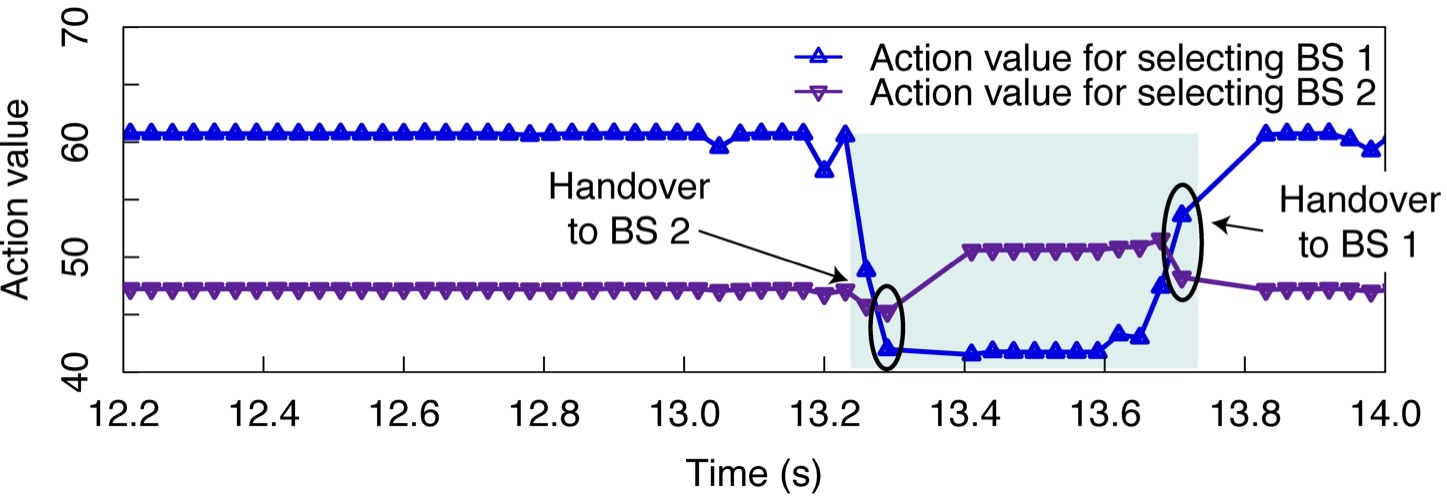}}	
	\caption{Comparison between time series of achievable data rate provided by BS~1 and that of learned action values. The action value, i.e., the estimation of the cumulative sum of the future data rates, decreases after the data rate decreases, which indicates that the degradation in the data rate provided by BS~1 cannot be predicted in a proactive manner.}
	\label{fig:action-value_rss}
\end{figure}

\subsubsection{Results}

In this experiment, it was shown that the obstacle-caused degradation in the data rates could not be predicted in a proactive manner by analyzing the learned action-value function in Fig.~\ref{fig:action-value_rss}. 
Fig.~\ref{fig:action-value_rss} shows the time series of the achievable data rate provided by BS~1 and the corresponding learned action values.
The data rate provided by BS~1 is degraded from approximately 13.25\,s to 13.70\,s because a pedestrian walks between the measurement device and transmitter.
First, we observe in Fig.~\ref{fig:action-value_rss} (a) that the data rate oscillates within approximately 90\,ms before the degradation occurs, and thus, the time-variation in the received powers is successfully observed before the degradation, as confirmed in other propagation experiments\cite{double_knife, flexible, koda_measurement2, 11ad_channel}.
However, the action value decreases sharply after the degradation in the data rate provided by BS~1.
As the action value is defined as the expected sum of the future performance, we can conclude that the obstacle-caused degradation of data rates in a mmWave link cannot necessarily be predicted proactively based only on the variation in the received powers.

\begin{figure}[t!]
	\centering
	\includegraphics[width = 0.93\columnwidth]{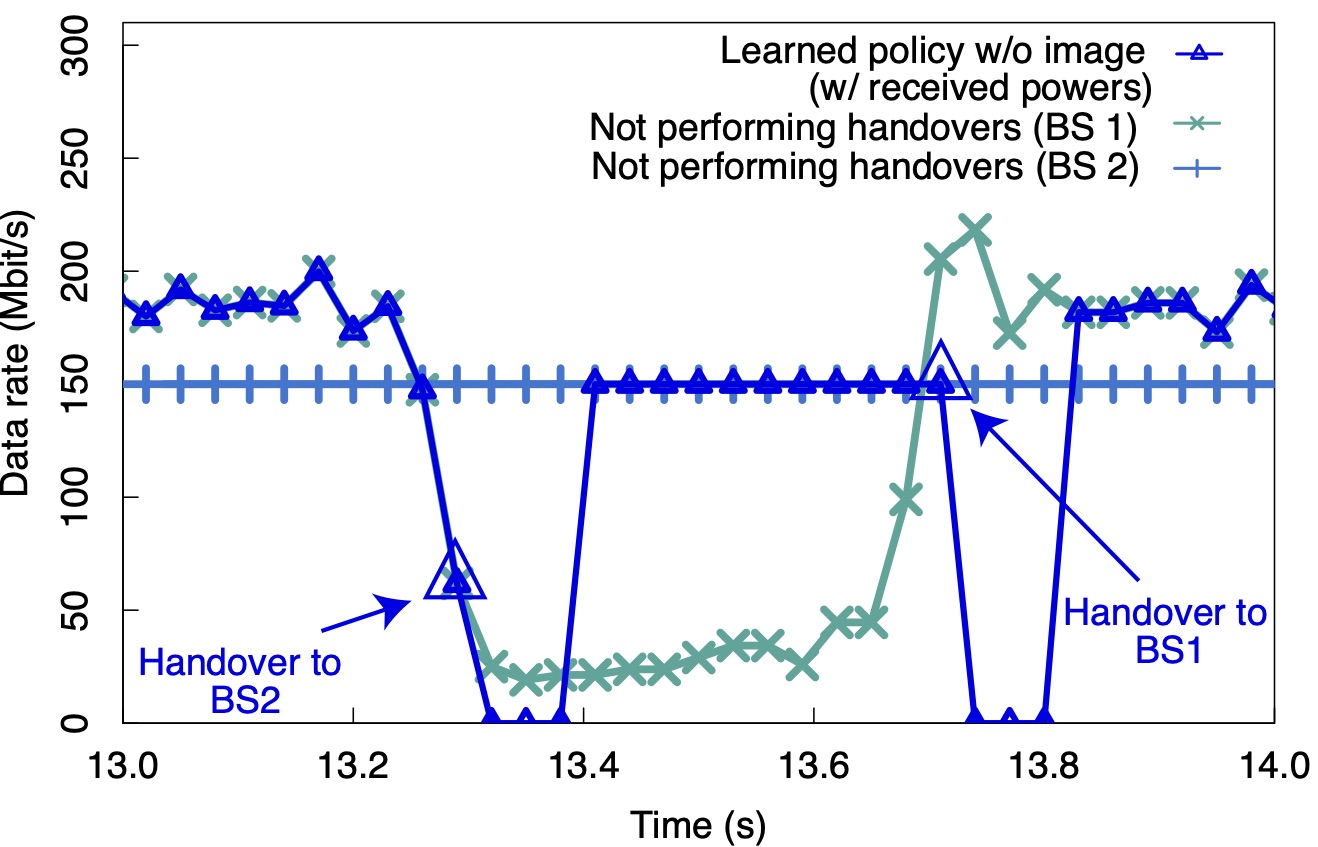}
	\caption{Example of handover timing when $T_{\mathrm{dis}} = 0.09$\,s.
		Handovers are performed after the variation in the data rate provided by BS~1.}
	\label{fig:rate_time_009_rss}
\end{figure}

Owing to the characteristics of the action-value functions, handovers are performed after the variation in the data rate provided by BS~1. 
Fig.~\ref{fig:rate_time_009_rss} shows an example of the time-varying data rate provided by our image-based handover framework when the service disruption time $T_{\mathrm{dis}} = 0.09$\,s.
It can be observed that a handover is performed after the variation in the data rate, the degradation is experienced within approximately 60\,ms.
If the handover is performed earlier, we can prevent the occurrence of the degradation in the data rate provided by BS~1 and enhance the time-average of the data rates.

It should be noted that, if we observe the received powers to have a short time-resolution, e.g., one millisecond, we could predict the degradation in the data rate provided by BS~1 from based on the time-variation in the received powers that occurred before the degradation.
However, as the time-variation occurred within approximately 90\,ms, the  degradation cannot be predicted from several hundreds of milliseconds before the degradation.
This example motivated us to develop a framework using other state information that exhibits more informative features for predicting even such degradation in the data rates in a proactive manner.

\section{Image-Based Handover Framework}
\label{sec:image_based}

This section details a proactive framework wherein  the handover timings are decided while the future degradation in the data rates is predicted in a proactive manner.
First, to enable the proactive prediction, we expand the state information such that the state includes time-consecutive camera images.
Using the time-consecutive camera images, we can capture the spatiotemporal dynamics of obstacles that are informative for predicting the degradation.
We then demonstrate that with the expansion of the state space, the degradation can be predicted from several hundreds of milliseconds in advance and confirm that a performance gain is realized owing to the proactive prediction.

\subsection{State Space Expansion for Proactive Prediction}
	For the network controller to leverage camera images for making handover decisions, we expand the state space in the previous section such that the state includes consecutive camera images.
	Let the number of time-consecutive camera images used in making handover decisions be denoted by $N$.
	We set the state space as follows:
		  \begin{align}
			\label{eq:state_set_img}
		      \mathcal{S}_{\mathrm{img}}\coloneqq \mathcal{S}_{\mathrm{rp}}\times \underbrace{\mathcal{X}\times\dots\times\mathcal{X}}_{N},
		  \end{align}
	where $\mathcal{X}$ denotes the set of all possible images.
	It should be noted that we consider the same actions, rewards, and state transition rules as \eqref{eq:action_set}--\eqref{eq:c_t_transition}, respectively, to obtain a  fair comparison of the performances achieved with the state space $\mathcal{S}_{\mathrm{rp}}$ and $\mathcal{S}_{\mathrm{img}}$, respectively.

	The state design enables an RL to predict the future data rate degradations in mmWave links caused by moving obstacles and facilitates the maximization of the expected cumulative sum of the future data rates as in \eqref{eq:optimal_policy}.
	This is because the state involving time-consecutive camera images reflects the spatiotemporal dynamics of the moving obstacles---for example, the dynamics of the obstacles approaching a LOS path---thus, reflecting the behavior of the data rates provided by deployed BSs at the future decision epochs $t + 1, t + 2, \dots$, which may comprise the decision epochs, in which one of the BSs is blocked.
	We demonstrate that the novel state design allows us to predict the degradation in the data rates caused by moving obstacles from several hundreds of milliseconds before the degradation occurs in the following section.

	In the evaluation, we do not use the time-series of the received power values in the state information because the information of the received power values may be redundant in the case of the existence of consecutive camera images.
	This is because the camera images $x_{t}, x_{t - 1}, \dots, x_{t - N + 1}$, reflect the spatial features at the decision epochs $t, t - 1, \dots, t - N + 1$, which may also be informative for capturing the received power values $p_{t}, p_{t - 1}, \dots, p_{t - N + 1}$ because the received power values are heavily dependent on the spatial features, such as the distance between a transmitter and receiver and the positions and shapes of obstacles that obstruct the path between the receiver and transmitter\cite{double_knife}.
	Thus, in this evaluation, we consider the state space as follows:
	\begin{align*}
		\label{eq:state_set_img2}
		\hat{\mathcal{S}}_{\mathrm{img}}\coloneqq \underbrace{\mathcal{X}\times\dots\times\mathcal{X}}_{N}\times\mathcal{J}\times\mathcal{C}.
	\end{align*}
	That is, we omit the sets of received power values $\mathcal{P}$ from the state space $\mathcal{S}_{\mathrm{img}}$.

\subsection{Experimental Evaluation}
\label{subsec:experiment_img}
We evaluate the image-based handover framework discussed above. 
The objective of this evaluation is to verify the feasibility of the proactive prediction on data rate variation caused by moving obstacles if a camera is available for performing the prediction.
Hence, further issues incurred by introducing cameras, such as costs for camera installments, were left aside, and we perform the evaluation focusing on the objective.
In the next section, we detail the evaluated scenario, and we provide realistic scenarios where the results from this feasibility study can be applied possibly without any additional costs for camera installments.
\begin{figure}
	\centering
	\includegraphics[width=\columnwidth]{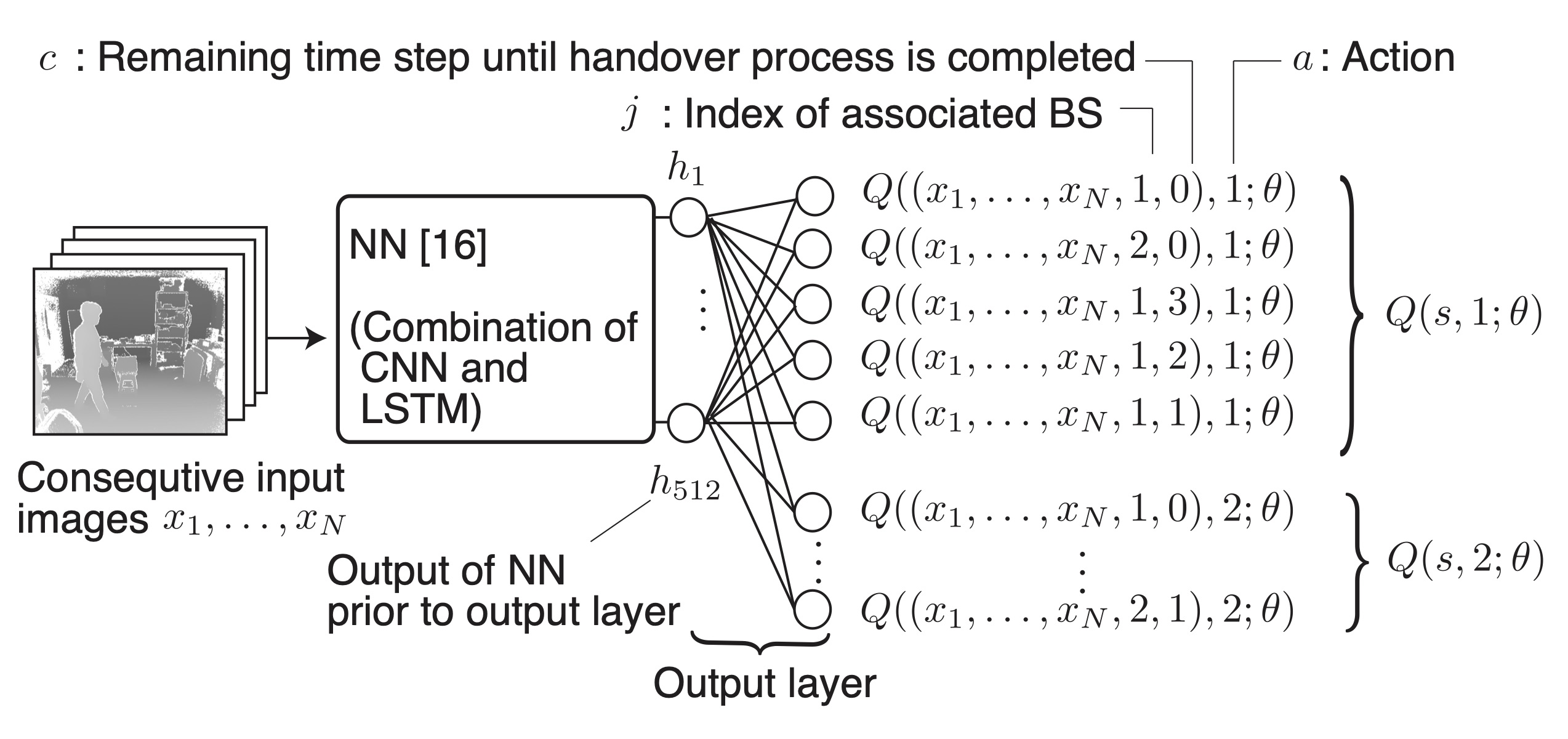}
	\caption{NN architecture for approximating optimal action-value function $Q^{\star}(s, a)$ defined in \eqref{eq:opt_q_func} for $\mathcal{C} = \{0,1, 2,3\}$ and $\mathcal{J} = \{1, 2\}$.
	With the exception of the output layer, the architecture herein is identical to that used in \cite{okamoto_access}.
		The architecture is a combination of a CNN, which deals with images,
		and an LSTM, which deals with sequential inputs\cite{goodfellow}.
	}
	\label{fig:nn_arc}
\end{figure}

\begin{figure*}[t!]
	\centering
	\includegraphics[width = 0.8\textwidth]{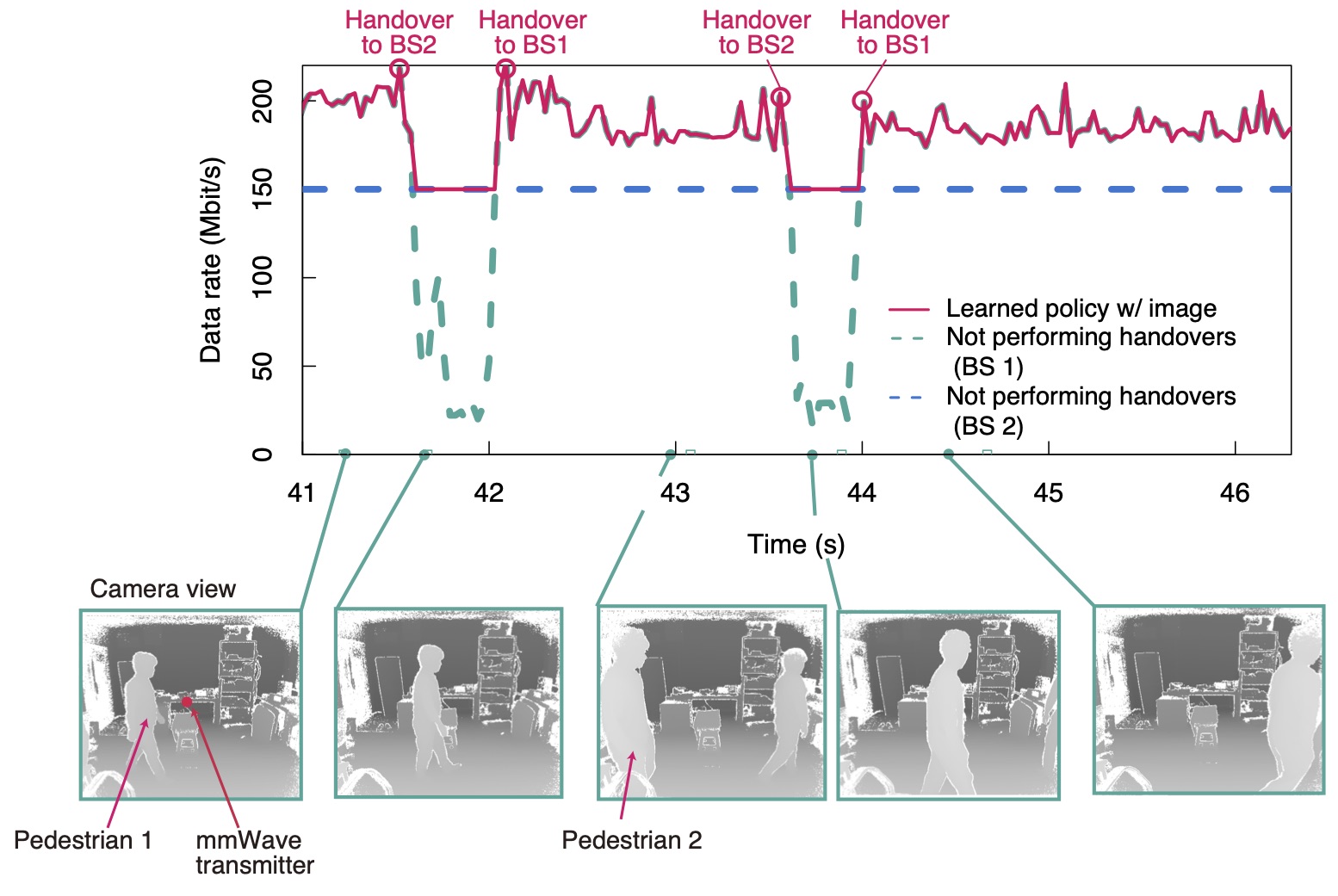}
	\caption{Time series of the achievable data rates under the condition of the service disruption time of $T_{\mathrm{dis}} = 0$\,s and the corresponding camera images.}
	\label{fig:rate_time}
\end{figure*}
\subsubsection{Evaluated Scenario}
\label{subsubsec:evaluated_scenario2}
We consider the scenario as discussed in Section~\ref{secV:subsec:evaluated_scenario} with regard to the deployment of the BSs and STA, channel, initial access procedure, and the coverage area of the BSs.
In the scenario, a camera monitors the two pedestrians that walk between BS~1 and the STA.
As we have assumed that BS~2 is free of blockages, we do not perform the proactive prediction in the performance of the link between BS~2 and the STA.

This experiment is performed by fixing measurement device and transmitter positions and changing camera angles motivated by the objective of this evaluation.
	The objective of this experiment is to validate the feasibility of the proactive prediction achieved by introducing cameras in the two basic angles.
	Hence, it is beyond the scope of this study to perform experiments in various configurations such as in terms of the parameters irrelevant to cameras.

There are some realistic scenarios to which the results from this feasibility study can be applied.
	In this feasibility study, it is examined that we can at least perform the proactive prediction with camera images if an STA and BSs are static, and the order of the distances from the STA and BSs are several meters.
	Hence, we can expect that the results are also applied to, for example, a video streaming to static wireless monitors in an office, where the STA and BSs are also static and the order of the distances from an STA to BSs are several meters.

Moreover, in such realistic scenarios, additional costs are not necessarily incurred when we can utilize pre-installed cameras.
In concrete, given the aforementioned scenarios transmitting streaming data, we can utilize pre-installed surveillance cameras monitoring the entire office. 
In this case, the results from this feasibility study can be applied possibly without any additional costs for camera installments.

Concerning a real implementation, we evaluate the image-based handover framework in the two totally different camera angles shown in Fig.~2 (left).
The camera angles affect how the obtained images represent the movement of pedestrians, and thus, they may also affect the feasibility of the proactive prediction more strongly compared to other parameters irrelevant of cameras such as the distance of the STA and BS or their heights.
Hence, concerning the camera angles may be important for a real implementation, and we perform the evaluation in the two basic camera angles that are orthogonal to each other.

\subsubsection{Procedure of Performing Decision Process}
We perform the decision process in the image-based handover framework using a procedure similar to that used in the previous evaluation with the exception that the state includes consecutive camera images obtained in the experiment.
Let $x_t$ denote the camera image (that contains 40$\times$40 pixel values in the experiment and obtained with the frame rate of 30 frame per second) obtained at the same time when the received power value $p_{1, t}$ is obtained.
From the state definition in \eqref{eq:state_set_img2}, we replace the received power values $p_{t}, p_{t - 1}, \dots, p_{t - N + 1}$ in the state $s_t$ in the previous evaluation with the time consecutive images $x_{t}, x_{t - 1}, \dots, x_{t - N + 1}$.
We learn the optimal policy using deep RL with an NN that is specifically used for handling the time consecutive camera images as discussed in the following section.
The parameter associated with the deep RL is set as shown in Table~\ref{table:param}.

\subsubsection{Neural Network Architecture}
\label{subsec:NN_arc}
In the deep RL, an NN is trained such that the NN is a good approximation of the optimal action-value function $Q^{\ast}(s, a)$ in \eqref{eq:opt_q_func}\cite{DQN}. 
We focus on the NN architecture designed to perform deep RL in the decision process discussed in the previous subsection\footnote{The NN is trained using the method discussed in \cite{DQN}.
For details of the training, \cite{DQN} may be referred to.}.

We design the NN architecture such that the NN has separate outputs for each possible combination of $j\in\mathcal{J}$, $c\in\mathcal{C}$, and $a\in \mathcal{A}$, as shown in Fig.~\ref{fig:nn_arc}.
The design allows us to divide the parameters into two parts: the parameters associated with the camera images and those associated with the other low-dimensional observations $j$, $c$, and $a$.
Let $Q(s, a;\theta)$ be the NN, where $s\in\hat{\mathcal{S}}_{\mathrm{img}}$, $(x_1,\dots, x_N)\in\mathcal{X}^N$, and $\theta$ denote the parameters of the NN.
In the architecture, the NN is expressed as follows:
\begin{align}
	Q\bigl((x_1,\dots, x_N, j, c), a;\theta\bigr) = \sum_{k = 1}^{512}\theta_{j, c, a, k}h_{k},
\end{align}
where $h_{1},\dots,h_{512}$ denote the output values of the layer prior to the output layer and $\theta_{j, c, a, 1},\dots,\theta_{j, c, a, 512}$ denote the parameters in the output layer corresponding to the combination of $j$, $c$, and $a$.
The parameters used to obtain the output values $h_{1},\dots,h_{512}$ are associated with the camera images, and the parameters in the output layer, $\theta_{j, c, a, 1},\dots,\theta_{j, c, a, 512}$ are associated with the low-dimensional observations $j$, $c$, and $a$.

The motivation for the architecture is that it is necessary to use the observations $j$ and $c$ for handover control.
In our problem setting, the state $s$ consists of $N$ consecutive images $(x_1,\dots, x_N)$ with thousands of elements and $(j, c)$ with only two elements.
If we let the input of the NN be $(x_1,\dots, x_N, j, c)$, and thereby, process the camera images $(x_1,\dots, x_N)$ and $(j, c)$ using the same parameters, the variation in $(j, c)$ does not significantly impact the NN output values.
This is because NNs generally estimate the feature representations of overall inputs; thus, they do not propagate the variation in one or two elements in the inputs to the output\cite{goodfellow}.
Hence, the controller can ignore the variation in $(j, c)$ while making a handover decision.

It should be noted that we employ the NN architecture used in \cite{okamoto_access} with the exception of the output layer.
The architecture is reported to facilitate the prediction of a future data rate in an mmWave link based on camera images.
Hence, it is expected that the architecture also facilitates the learning of the optimal action-value function, which is the cumulative sum of the performance data rates in our problem setting.

\subsubsection{Results}
\label{subsec:results}

\begin{figure*}[t]
	\centering
	\subfigure[Time series of the data rate provided by BS~1 obtained when the camera monitors with angle~A.]{\includegraphics[width = \columnwidth]{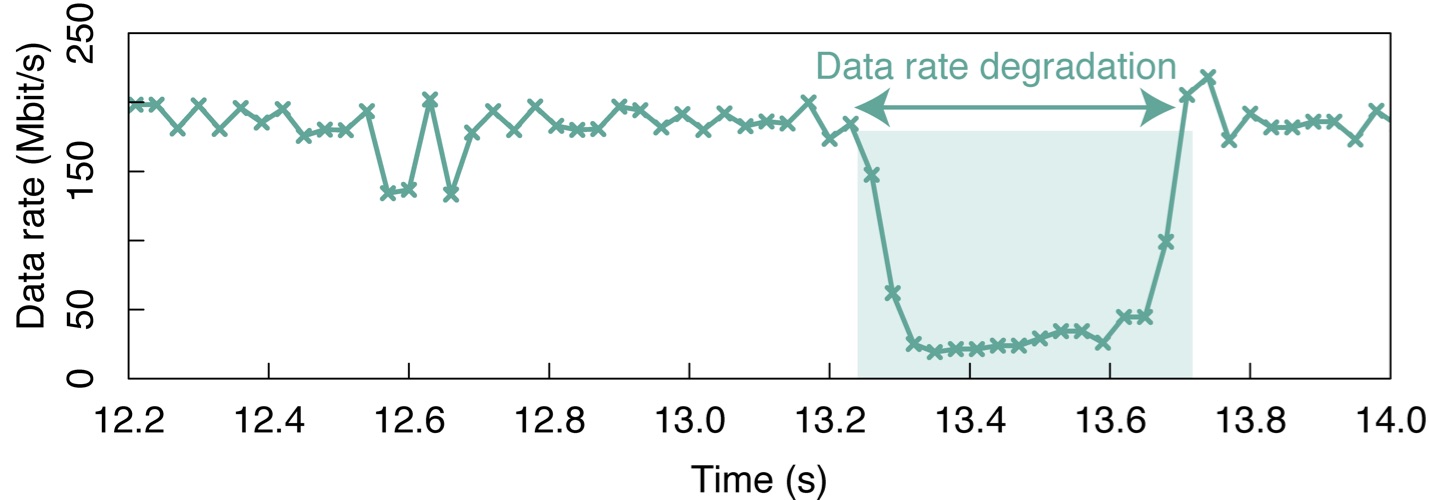}}\hspace{1em}
	\subfigure[Time series of the data rate provided by BS~1 obtained when the camera monitors with angle~B.]{\includegraphics[width = 0.9\columnwidth]{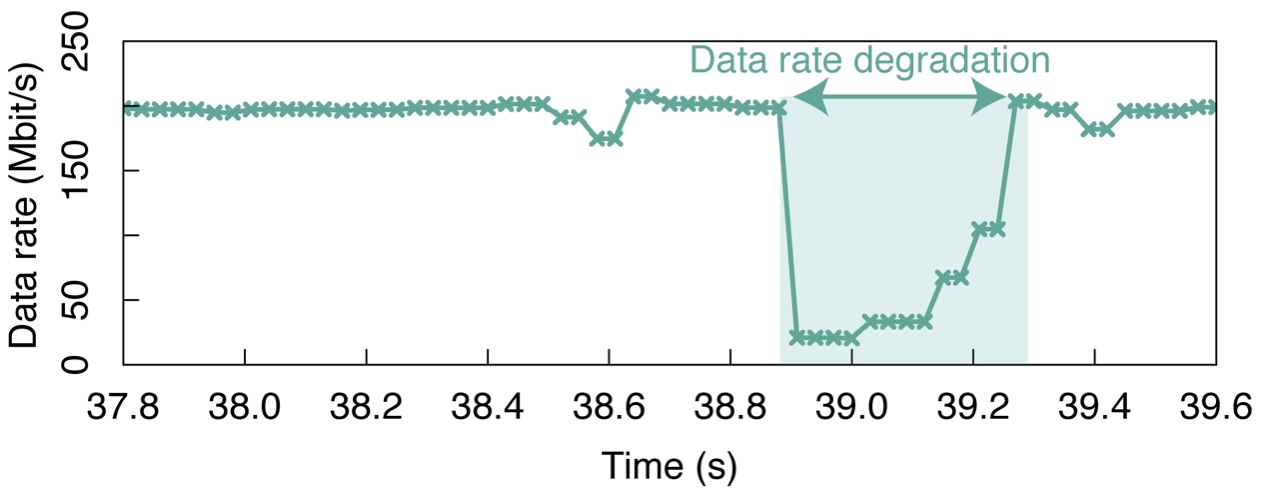}}
	\subfigure[Time series of action value in proposed framework when the camera monitors with angle~A.]{\includegraphics[width = \columnwidth]{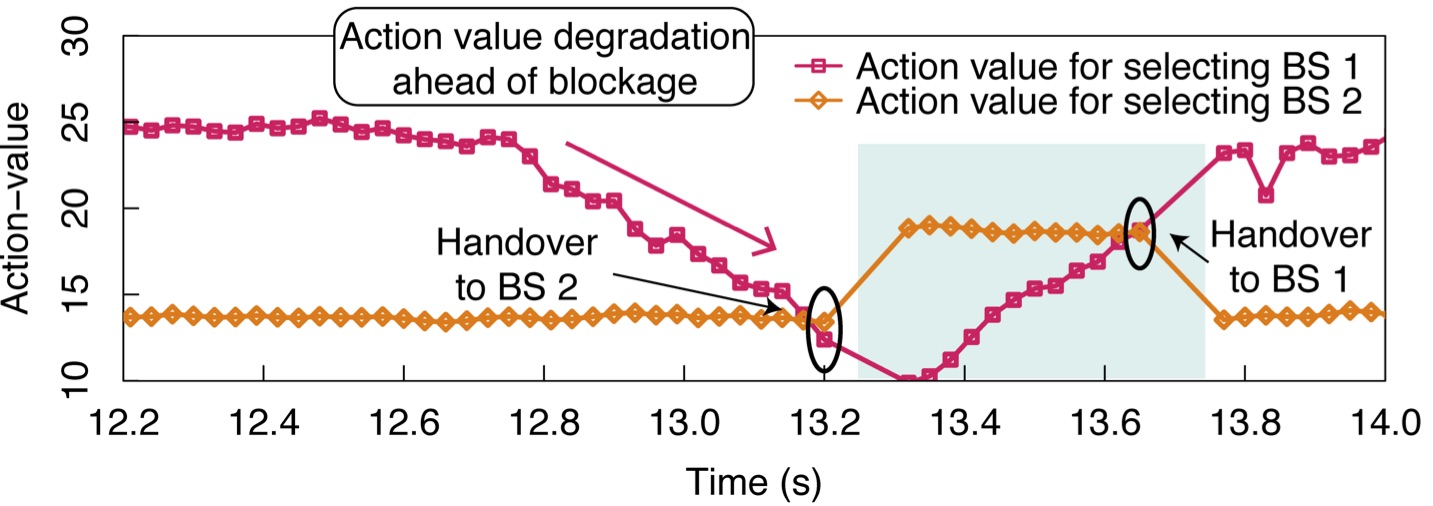}}\hspace{1em}
	\subfigure[Time series of action value in proposed framework when the camera monitors with angle~B.]{\includegraphics[width = 0.9\columnwidth]{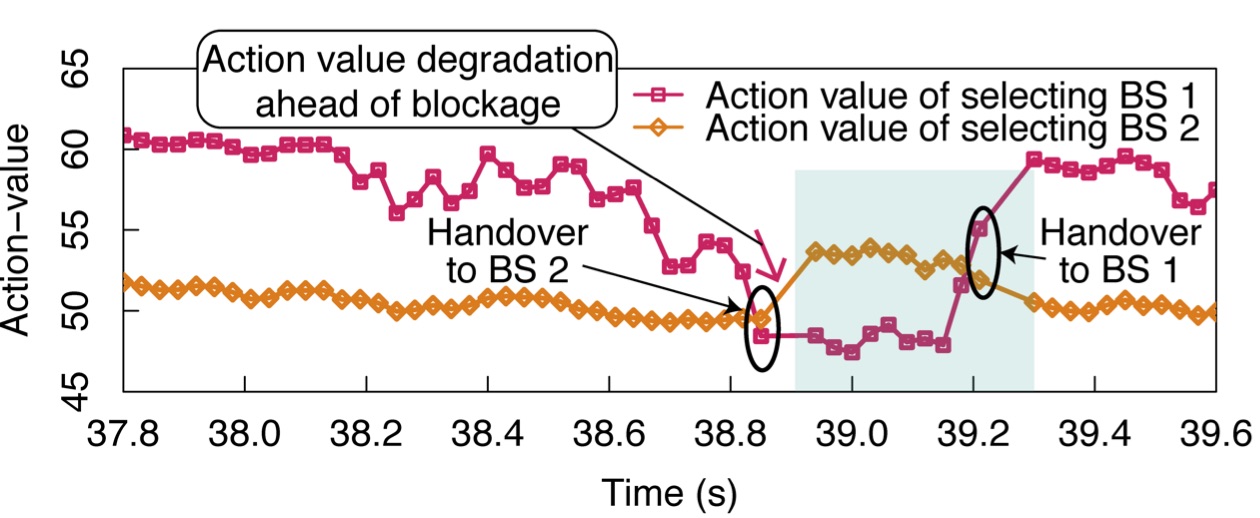}}
	\caption{Comparison between time series of the data rate provided by BS~1 and that of the learned action-value function. The action value in the proposed image-based framework decreases several milliseconds before the performance degradation at BS~1, which indicates that the proposed framework successfully predicts the future performance degradation in advance (the action value is defined as the expected cumulative discounted sum of the future performance).}
	\label{fig:action-value}
\end{figure*}

\begin{figure*}[t!]
	\centering
	\subfigure[Portion of the time series of the data rate and hanover timing learned when the camera monitors with angle~A.]{\includegraphics[width = 0.93\columnwidth]{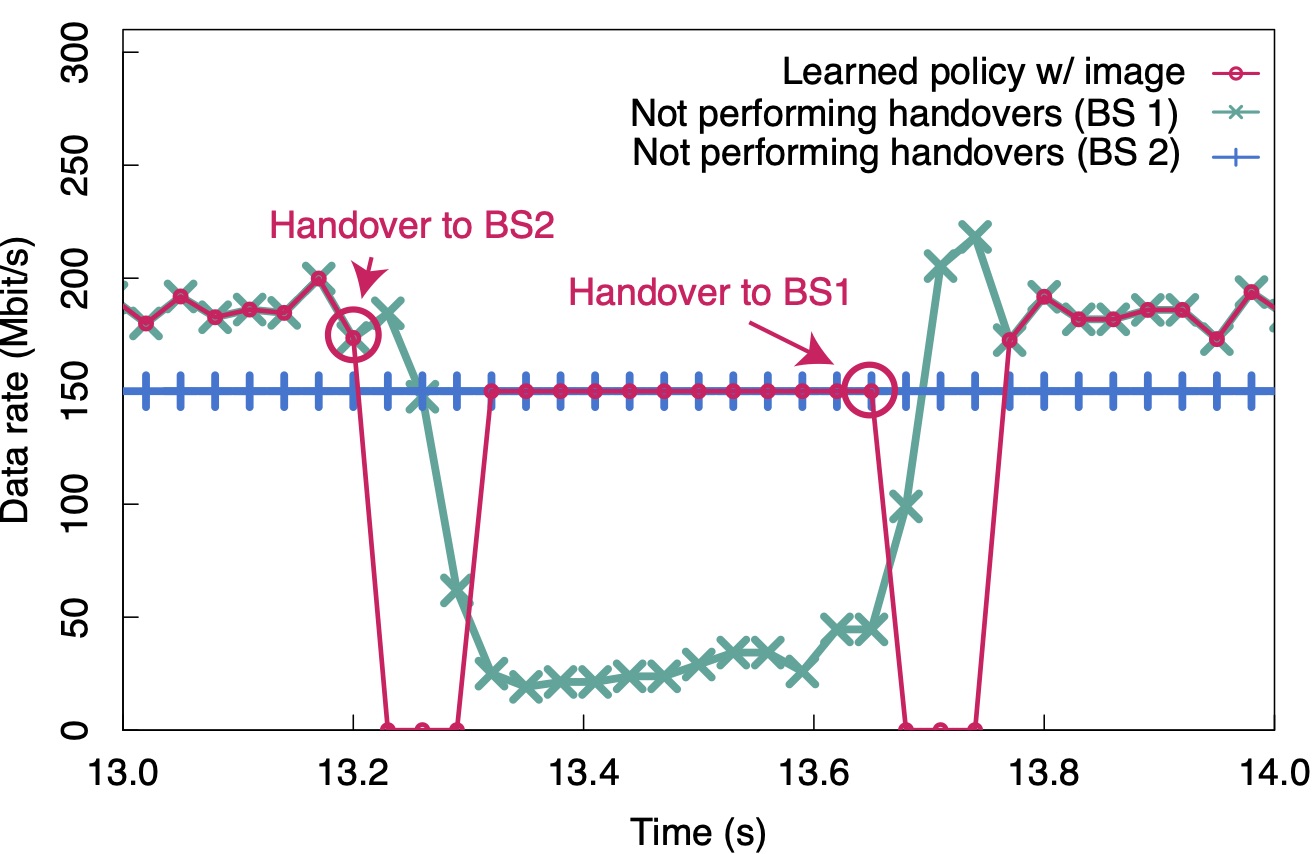}}\hspace{1em}
	\subfigure[Portion of the time series of the data rate and hanover timing learned when the camera monitors with angle~B.]{\includegraphics[width = 0.89\columnwidth]{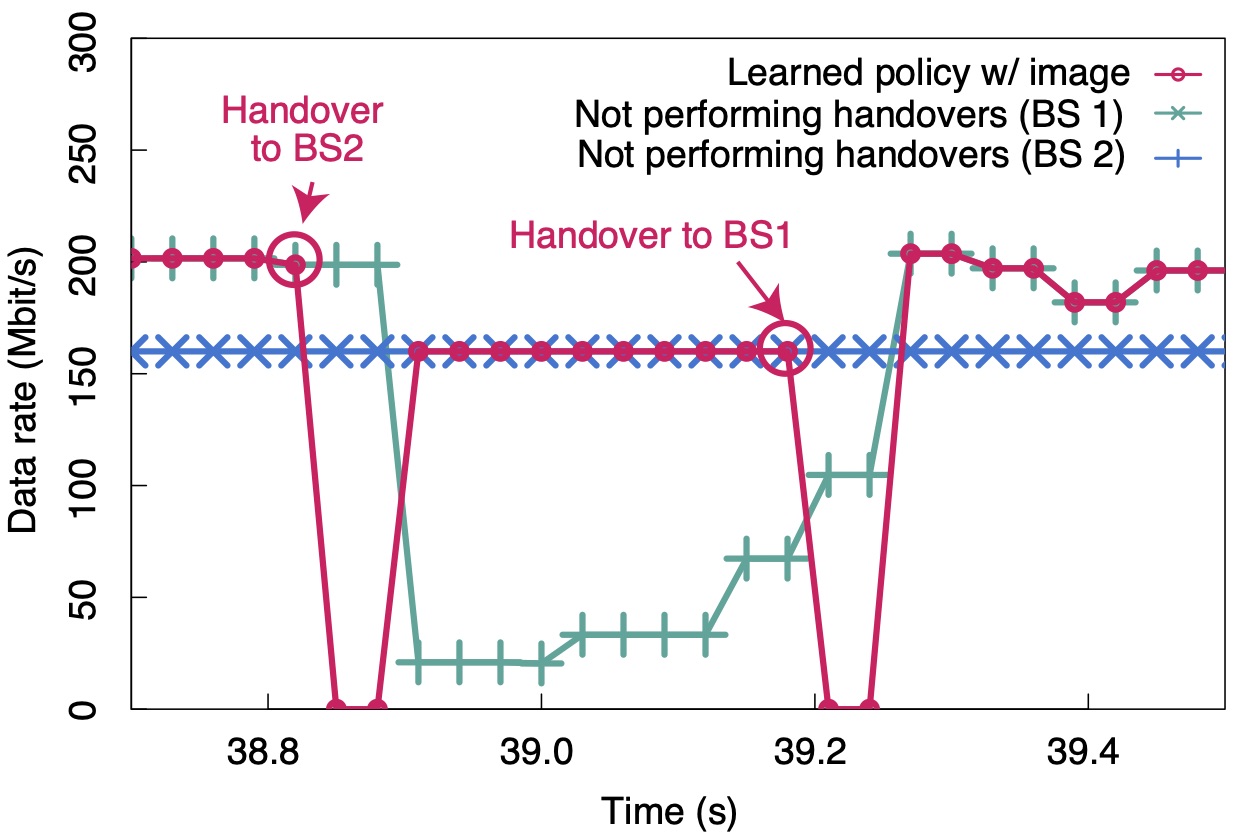}}
	\caption{Example of handover timing.
		The proposed image-based framework performed handovers before the variation in the channel at BS~1 and STA with either camera angle~A or angle~B.}
	\label{fig:rate_time_009}
\end{figure*}

We confirm that the deep RL successfully maximizes the time-average of the achievable data rate in the mmWave links in the state design in \eqref{eq:state_set_img2}.
Fig.~\ref{fig:rate_time} shows an example of a time series of the data rate in the case wherein $T_{\mathrm{dis}} = 0$\,s.
The pedestrians walk in front of the mmWave transmitter at approximately 41.5\,s and 43.9\,s.
At the same time, the data rate provided by BS~1 is degraded from approximately 200\,Mbit/s to 30\,Mbit/s.
Our framework successfully selects the BS that provides a higher data rate than the counterpart at each decision epoch and thereby maximizes the overall data rate.

It should be noted that in Fig.~\ref{fig:rate_time}, the intervals between the two successive handovers, i.e., the handover from BS~1 to BS~2 and that from BS~2 to BS~1 are according to the durations wherein the blockage effects sustained.
This results can be interpreted that our image-based handover framework can form a handover strategy while predicting such durations wherein the blockage effects sustain in an end-to-end manner.
In this regard, we've achieved the prediction of such durations implicitly in learning the handover policy.

We show that the proposed framework predicts a future data rate degradation from several hundreds milliseconds before the degradation occurs by analyzing the learned action-value function shown in Fig.~\ref{fig:action-value}.
Fig.~\ref{fig:action-value} shows the learned action value at each decision epoch before and after the blockage effect in Fig.~\ref{fig:rate_time_009_rss}.
We can see that the action value begins to decrease from approximately 500\,ms (in camera angle~A) or 200\,ms (in camera angle~B) before the actual degradation in the data rate provided by BS~1.
As the action value is defined as the expected sum of the future data rates, we can consider that our image-based framework successfully predicts the future performance degradation several hundred milliseconds before the blockage effects occur.

Owing to the proactive prediction, our image-based handover framework triggers a handover in a proactive manner.
Fig.~\ref{fig:rate_time_009} (a) shows an example of a time-varying data rate provided by our image-based handover framework with  camera angle~A.
The plotted duration corresponds to that in Fig.~\ref{fig:rate_time_009_rss}.
Our proposed framework is different from the received-power-based prediction presented in the previous section in Fig.~\ref{fig:rate_time_009_rss} and successfully triggers handovers prior to the variation in the data rate provided by BS~1.
Fig.~\ref{fig:rate_time_009}(b) shows an example of a time-varying data rate provided by the proposed image-based handover framework with the camera angle~B.
Similarly, with a different angle, the image-based framework triggers handovers prior to the degradation of the data rate provided by BS~1.

It should be noted that results show the feasibility of the proactive prediction even with the image time-series with 40$\times$40 pixels and a frame rate of 30 frames per second.
Hence, to accomplish the proactive prediction, it is sufficient to leverage such qualities of image videos, which cannot be obtained only with sophisticated cameras exampled by Kinect sensors but also with commercial products of smart phones\cite{ji2014vehicular} or surveillance cameras\cite{hill2009measuring}.

\begin{figure}[t]
	\centering
	\includegraphics[width = 0.8\columnwidth]{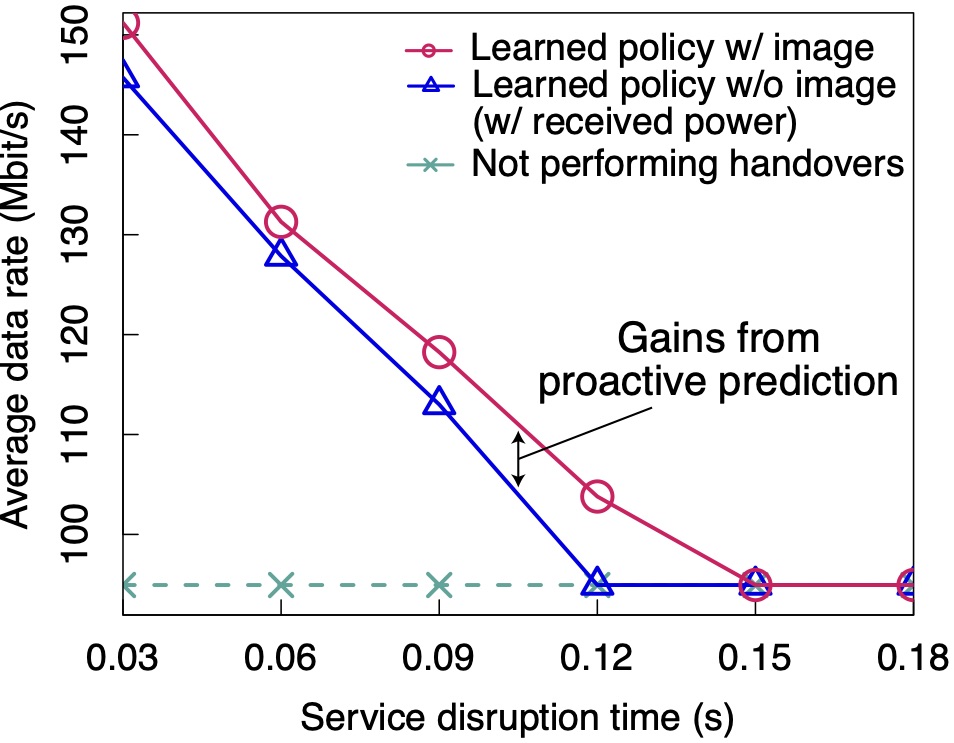}
	\caption{Performance comparison between proposed image-based framework and received power-based framework under various service disruption times $T_{\mathrm{dis}}$ when the camera in angle~A.}
	\label{fig:result}
\end{figure}

\begin{figure}[t]
	\centering
	\includegraphics[width = 0.8\columnwidth]{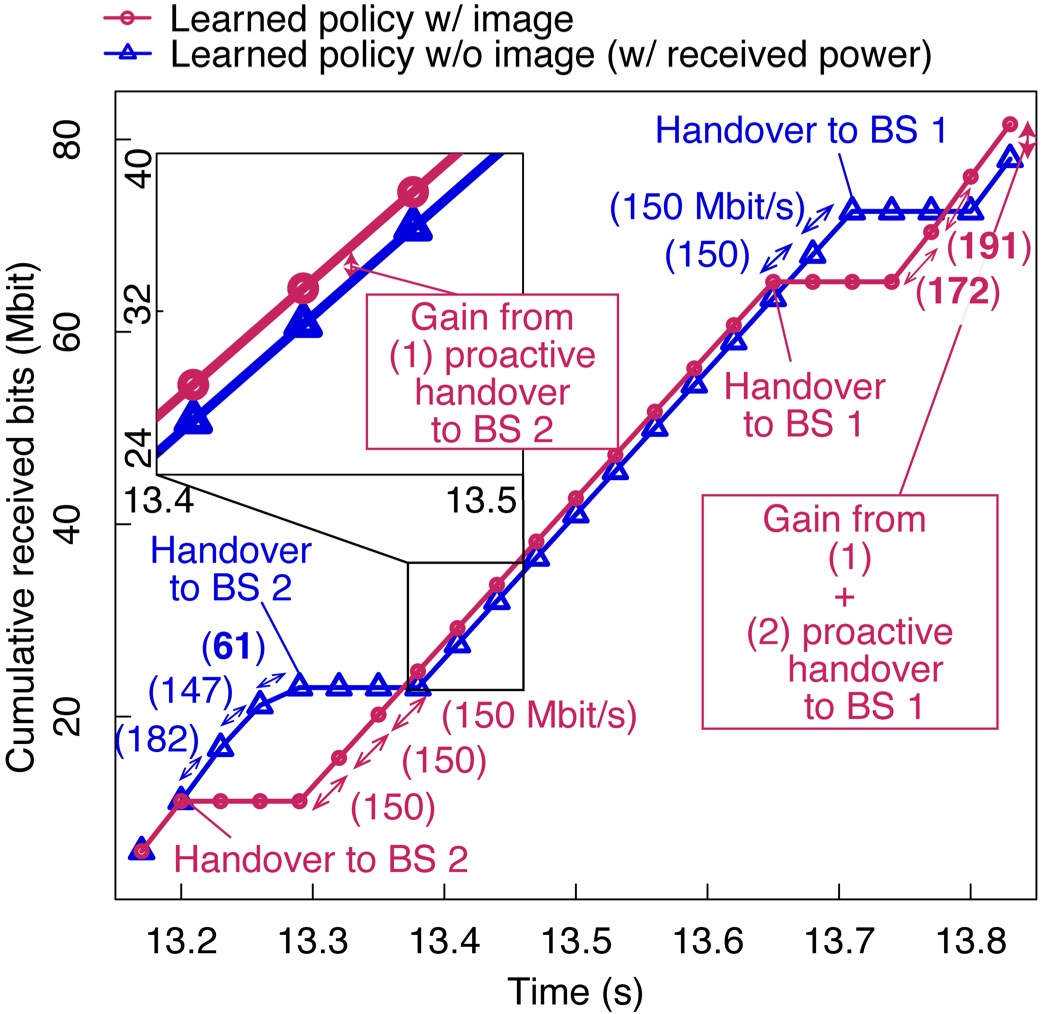}
	\caption{Comparison of cumulative received bits between our image-based framework and the received-power-based framework.}
	\label{fig:cumulative_gain}
\end{figure}

We compare the proposed image-based handover framework with a handover framework that does not leverage images, i.e., the received power-based handover framework.
Fig.~\ref{fig:result} shows the average data rate achieved by the two handover policies over the duration of specific events 200\,ms before and after a blockage\footnote{In detail, the blockage event is defined as the event where the received power is 3\,dB below from the one observed in a LOS condition, which is according to \cite{choi2018measurement}.}.
As blockage event, the one depicted in Figs.~5 and 9(a) has been chosen since handover policies with and without camera images exhibit different behavior according to the aforementioned figures.
The choice of 200\,ms is attributed to the fact that the two handover policies exhibited different a behavior from at most 200\,ms before and after the blockage event.
From Fig.~\ref{fig:result}, we can see that the handover policy learned with images achieves a higher or equal data rate as compared to the policy learned without images.

A realistic scenario where we can benefit from the gain is exemplified by combining Agile-Link\cite{hassanieh2018fast} as the beam search method and make-before-break\cite{gimenez2017towards} as the handover procedure.
In such a scenario, the service interruption subjected by a beam alignment is under 1\,ms with a 128 size array, and by the other handover procedure would be tens of milliseconds.
This leads to an overall service interruption time $T_{\mathrm{dis}}$ of several tens milliseconds.
Recalling that there is a gain from the proactive handover when $T_{\mathrm{dis}}$ is in the order of tens milliseconds in Fig.~10, the system benefits from the gain in such a scenario.

To illustrate how the proactive handover led to the performance gain provided in Fig.~\ref{fig:result}, we show the cumulative received bits in the proposed image-based handover framework and in the received power-based framework.
	Fig.~\ref{fig:cumulative_gain} shows the amount of cumulative received bits from the time 200\,ms before the blockage event.
	The horizontal axis corresponds to that either in Figs.~5 and 9(a).
	After a handover to BS~2 is performed in the image-based handover framework, the amount of cumulative data bits is temporarily lower than that in the received power-based handover framework.
	Meanwhile, the amount of cumulative received bits in the image-based handover framework is larger than that in the power-based framework by 1.7\,Mbit from instant 13.4\,s.
	These results confirm the benefits of proactive handover in the long run, and the increase of the received bits can be interpreted as the gain from proactively performing a handover to BS~2.
	Similarly, the amount of cumulative bits in the image-based handover framework is larger than that in received power-based framework by 3.6\,Mbit from instant 13.8\,s. 
	This can be attributed to the fact the image-based framework benefits earlier from a recovering data rate in BS~1 while the STA remains to be associated with BS~2 in the received power-based framework.
	The increase in received bits can also be interpreted as a gain from proactively performing handover to BS~1.

	\begin{figure}[t]
		\centering
		\includegraphics[width=0.8\columnwidth]{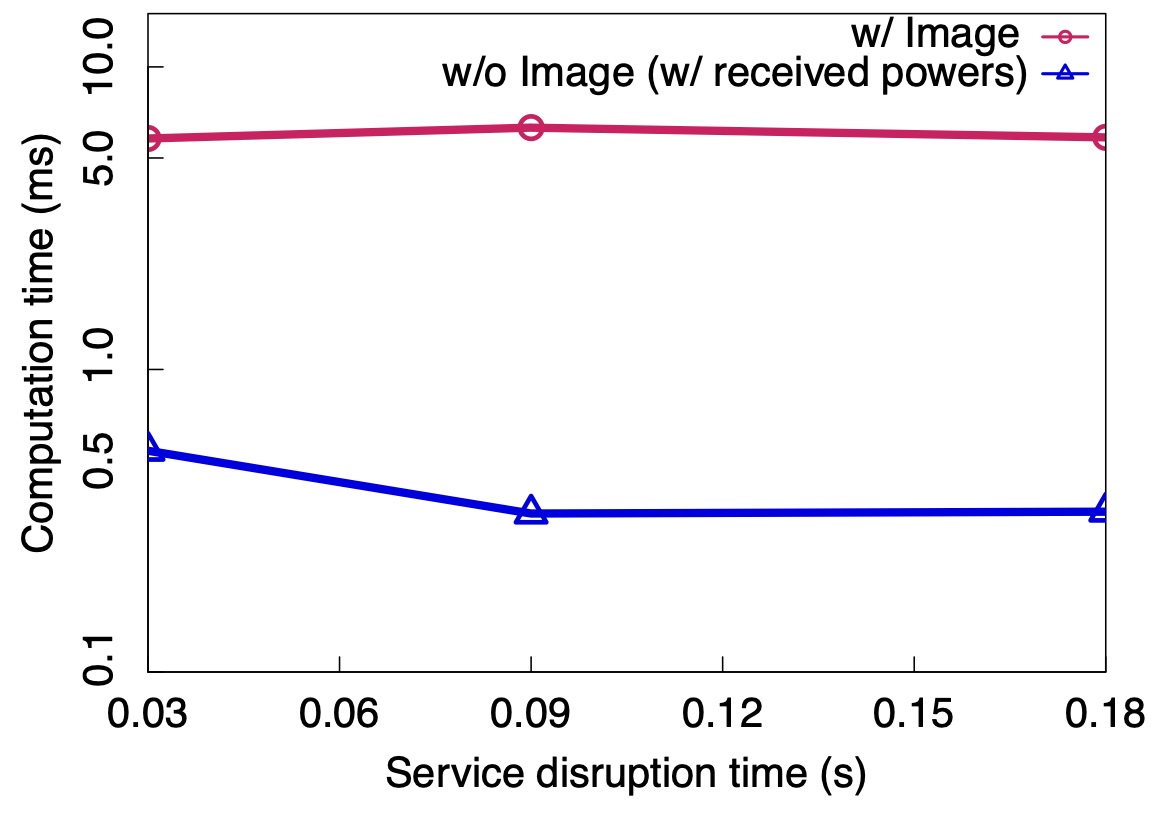}
		\caption{Computation time for making handover decisions.}
		\label{fig:computation_time}
	\end{figure}

We analyze the computation time required for making a handover decision in the context of an example.
	Fig.~\ref{fig:computation_time} shows an example of the computation time for making a handover decision.
	The computation time is defined as the time for calculating the action-value from an input of images and is measured with a GeForce GTX 1080 Ti GPU.
	The received power-based handover achieved the smaller computation time because of the lower dimensionality of the input.
	Meanwhile, in the proposed image-based handover framework, the computation time was still in the order of several milliseconds.
	The computation time is sufficiently shorter than the required handover interval, i.e., an interval between the two successive handovers, and is reported as 750\,ms in mmWave 5G systems\cite{talukdar2014handoff}.
	Thus, a shorter computation time relative to the required handover interval is possible.
	To this regard, the computation time incurred by the large-dimensionality of images can be overcome.

\begin{figure}[t]
	\centering
	\includegraphics[width=\columnwidth]{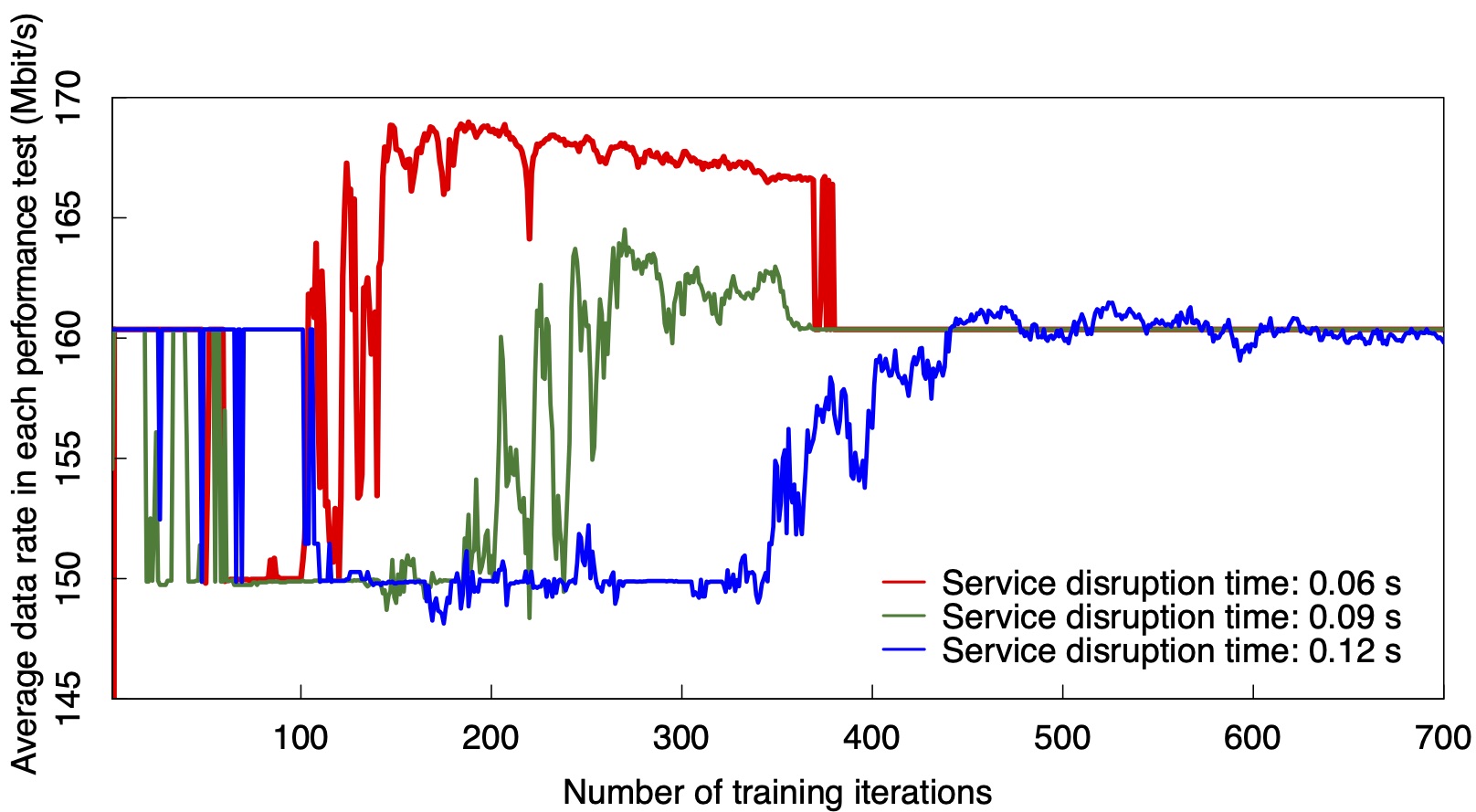}
	\caption{Average data rate in each performance test corresponding to a training iteration.}
	\label{fig:convergence}
\end{figure}

We investigate the convergence property of the training procedure in Fig.~\ref{fig:convergence}.
		Fig.~\ref{fig:convergence} shows the learning curve, i.e., the average data rate in the performance test corresponding to each training step.
		We obtained the following trends as the training steps are iterated: the performance enhancement, achievement of the maximum performance, and convergence to the degraded performance
		These results shows that the training procedure does not converge to maximum performance.
		This is attributed to the fact that the training process is not stable, which is commonly reported in deep RL\cite{DQN, sutton} when a non-linear approximator for the action-value function is used.
		These results motivate us to design an improved algorithm that converges to a maximum performance; however, seeking for the better convergence property is beyond the scope of this study.

Nevertheless, a practical solution can be employed to benefit from the results in this study.
	The solution named, ``off-policy evaluation framework''\cite{theocharous2015personalized}, keeps track of the best performing policy.
	As we are evaluating the policy that achieves maximum average data rates among the learned policies, we can benefit from the results in this study by designing the algorithm such that the off-policy evaluation framework is performed.

\section{Conclusion}
\label{sec:conc}
We presented a new paradigm for leveraging time-consecutive camera images in handover decision problems for realizing the proactive prediction of future long-term performances.
We first experimentally noted that the obstacle-caused data rate degradation in mmWave links cannot necessarily be predicted in a proactive manner based only on the time-variation of the received powers before the degradation. 
To solve this problem, we proposed the expansion of the state space in order for the state information to comprise consecutive camera images, which comprise informative features for proactively predicting long-term data rates in mmWave links.
To overcome the difficulty of the higher dimensionality of the expanded state space, we use deep RL for predicting the cumulative sum of the future data rates and deciding handover timings based on the predicted values.
By performing deep RL using the state information of experimentally obtained camera images, we confirmed that the state expansion allows the prediction of future obstacle-caused data rate degradation from approximately 500\,ms before the degradation occurs.
We also evaluated the time-average of the data rates over approximately two minutes and revealed that the proposed expansion of the state space resulted in a performance gain.

\bibliographystyle{IEEEtran}
\bibliography{IEEEabrv,main}

\end{document}